\documentclass{IEEEtran}
\usepackage[T1]{fontenc}
\UseRawInputEncoding
\usepackage{amsmath}
\usepackage{amsthm}
\usepackage{amssymb}
\usepackage{graphicx}
\usepackage[bookmarks=true,bookmarksnumbered=true,bookmarksopen=true,bookmarksopenlevel=1,
 breaklinks=false,pdfborder={0 0 0},pdfborderstyle={},backref=false,colorlinks=false]
 {hyperref}
\hypersetup{pdftitle={Your Title},
 pdfauthor={Your Name},
 pdfpagelayout=OneColumn, pdfnewwindow=true, pdfstartview=XYZ, plainpages=false}

\makeatletter
\theoremstyle{plain}
\newtheorem{lem}{\protect\lemmaname}
\theoremstyle{plain}
\newtheorem{prop}{\protect\propositionname}
\theoremstyle{remark}
\newtheorem{rem}{\protect\remarkname}
\theoremstyle{plain}
\newtheorem{cor}{\protect\corollaryname}

\IEEEoverridecommandlockouts
\usepackage{cite}
\usepackage{amsmath,amssymb,amsfonts}
\usepackage{algorithmic}
\usepackage{graphicx}
\usepackage{cuted}
\usepackage{amsmath,lipsum}
\usepackage{textcomp} 
\def\BibTeX{{\rm B\kern-.05em{\sc i\kern-.025em b}\kern-.08em
    T\kern-.1667em\lower.7ex\hbox{E}\kern-.125emX}}

\allowdisplaybreaks[4]

\makeatother

\providecommand{\corollaryname}{Corollary}
\providecommand{\lemmaname}{Lemma}
\providecommand{\propositionname}{Proposition}
\providecommand{\remarkname}{Remark}

\begin{document}
\title{Localization and Tracking for Cooperative Users in Multi-RIS-assisted
Systems: Theoretical Analysis and Principles of Interpretations}
\author{Peng Gao, Lixiang Lian, \IEEEmembership{Member, IEEE}, Yuan Shen,
\IEEEmembership{Senior Member, IEEE} \thanks{The material in this paper was presented in part at the IEEE International
Conference on Communications, Rome, Italy, 2023.}\thanks{Peng Gao is with the School of Information Science and Technology,
ShanghaiTech University, Shanghai 201210, China, and also with the
State Grid Electric Power Research Institute, NariARI Group Co., Ltd.,
Nanjing 211106, China. (e-mail: gaopeng8@sgepri.sgcc.com.cn).}\thanks{Lixiang Lian is with the School of Information Science and Technology,
ShanghaiTech University, Shanghai 201210, China. (e-mail: lianlx@shanghaitech.edu.cn).
(Corresponding author: Lixiang Lian).}\thanks{Yuan Shen is with the Department of Electronic Engineering, Tsinghua
University, Beijing 100084, China (email: shenyuan\_ee@tsinghua.edu.cn).}}
\maketitle
\begin{abstract}
Localization and tracking (LocTrack) are fundamental enablers for
a wide range of emerging applications. Reconfigurable intelligent
surfaces (RISs) have emerged as key components for enhancing the LocTrack
performance. This paper investigates a multi-RIS-assisted multi-user
(MRMU) LocTrack system, where multiple RISs collaboratively reflect
the position-bearing signals for information fusion at the base station,
leveraging spatial-temporal correlations in user positions. While
studies have shown these correlations improve localization accuracy,
their trade-offs with system complexity remain unclear. To address
this gap, we characterize the effectiveness of spatial-temporal correlation
priors (STPs) utilization in MRMU LocTrack systems using a metric,
termed efficiency of correlation (EoC). To further elucidate correlation
propagation and RIS interactions, we provide a \textquotedblleft correlation
information routing\textquotedblright{} interpretation of EoC through
random walk theory. EoC provides a principled performance evaluation
metric, that enables system designers to balance localization accuracy
enhancement against the increased complexity. Additionally, we investigate
the error propagation phenomenon, analyzing its convergence and asymptotic
behavior in MRMU LocTrack systems. Finally, we validate the theoretical
results through extensive numerical simulations.
\end{abstract}


\section{Introduction\protect\label{sec:Introduction}}

Localization and tracking (LocTrack) have been envisioned as key capabilities
for numerous emerging applications, such as integrated sensing and
communication (ISAC), internet of vehicles (IoV) and so on \cite{ISAC-loc0,ISAC-loc1,ISAC-loc2,IoV-loc1,NF-loc1,Loc-survey1,Loc-survey2,Loc-survey3}.
Global navigation satellite systems (GNSS), as the default solution
for outdoor LocTrack, are difficult to function effectively for line-of-sight
(LoS) obstructed area, e.g., urban area with dense buildings, underground
area and indoor scenarios\cite{Fund-YShen}. As a revolutionary technology,
reconfigurable intelligent surfaces (RISs) can act as extra anchors
and offer additional degrees of freedom (DoFs) for the electromagnetic
channel by artificially shaping the traveling signals. RISs have demonstrated
great potential for wireless LocTrack, especially for non-line-of-sight
(NLoS) scenarios\cite{RIS-surv}.

In RIS-assisted localization systems, analyzing performance limits
is crucial to bridge the gap between theory and practical implementation,
ensuring an efficient, robust, and reliable localization system. Unlike
conventional positioning systems assisted by multiple anchors, the
RIS-assisted positioning systems exhibit different characteristics
\cite{RIS-surv}. Firstly, the signal from the user arrives at the
base station (BS) through cascaded channel via RIS, embedding the
location information within the cascaded channel. Considering the
impact of multipath effects, extracting meaningful location information
from the cascaded channel becomes highly challenging. Secondly, high-accuracy
localization systems typically require multiple anchors, which can
introduce additional signaling and hardware overhead. In contrast,
RISs are cost-effective and flexible to serve as virtual anchors,
providing multiple DoFs while avoiding extra signal aggregation overhead.
Therefore, in multi-RIS-assisted systems, the BS naturally receives
signals aggregated from multiple RIS reflections. Third, RIS enhances
the quality of position-bearing signals through optimized beamforming,
improving the efficiency of position information extraction. Understanding
the localization performance in multi-RIS-assisted systems is crucial
for optimizing system design and implementation.

However, the existing literature on the multi-RIS-assisted localization
systems predominantly focuses on the LocTrack problem for single user
or multi-users without accounting for interactions among them\cite{Multi-RIS-wo-spatem1,Multi-RIS-wo-spatem2,Multi-RIS-wo-spatem3}.
In contrast, in many scenarios involving multi-user localization,
the states of multiple users, such as their motion directions or speeds,
often vary systematically following some predefined patterns. This
leads to discernible graphical structural features among the positions
of multiple users across both spatial and temporal domains, which
is referred to as the \emph{spatiotemporal correlations} of multi-users'
positions. For example, in scenarios such as a convoy of vehicles
traveling along a high-speed road, a swarm of unmanned aerial vehicles
(UAVs) flying towards a common destination to accomplish a task \cite{RIS-AoA3},
or a group of pedestrians walking along a sidewalk\cite{M_tar_tr_app2,M_tar_tr_app4},
various factors such as motion area limitations \cite{mot-cont-vehiclelane},
traffic regulations or underlying relationships among users impose
mutual restrictions on the movement states of multiple users\cite{Mot-constraint},
thereby revealing potential spatiotemporal correlations in their positions.
Such \emph{spatiotemporal correlation priors} (STPs) can be leveraged
in the design of LocTrack algorithms to enhance the accuracy in multi-RIS-assisted
multi-user (MRMU) LocTrack systems significantly.

Some of the emerging works have incorporated the spatial \cite{spatial-corr-2,spatial-corr-lian}
or temporal \cite{tRIS-tr1,RIS-tr2,RIS-tr2-lian} correlations of
user positions into the localization process. Even though the benefits
of correlation exploitation for localization accuracy have been extensively
verified empirically in \cite{Shen-Yuan-Preceeding,Shen-Yuan-JSAC},
whether the improvement is worth the additional system complexities,
especially in an MRMU LocTrack system, has not been considered. Moreover,
these results are limited to the performance of specific localization
algorithms and do not fully elucidate the fundamental relationship
between the final localization performance and spatiotemporal correlations
among users' positions. In MRMU LocTrack systems, the collaboration
among heterogeneous devices across both spatial and temporal domains,
along with the correlation priors in users' positions, plays crucial
roles in performance analysis. Understanding the role of prior information
in the LocTrack process, the efficiency of its utilization, and the
impact of multi-RIS cooperation presents fundamental yet challenging
performance insights. 

In large-scale networks, statistical probability models are commonly
employed to describe the spatial or temporal correlation of users'
positions. For example, Gamma-Markov model was introduced in \cite{spatial-corr-lian}
to capture the spatial correlation of vehicle positions in the platoon,
while Gauss-Markov model was employed in \cite{tRIS-tr1} to characterize
the temporal correlation of user positions in successive time steps.
To leverage the additional information provided by STPs, iterative
information exchanges between correlation priors, RISs, and measurements
collected at BS, as well as among different location states of users
are necessary \cite{tRIS-tr1,locTrak-Algor1,locTrak-Algor2}. Therefore,
the implementation of MRMU LocTrack algorithms entails high complexity,
which necessitates the consideration of whether the performance gains
obtained from the correlation priors justify the additional system
complexities.

Besides the additional system complexity, when utilizing the STP for
joint LocTrack in MRMU systems, the estimation processes of user positions
at different spatial-temporal instances are coupled together. The
estimation errors from other location states can affect the current
location state estimation, thus affecting the efficiency of correlation
prior utilization. This coupling is reflected in the fact that the
equivalent Fisher information matrix (EFIM) of user positions is not
block-diagonal but contains off-diagonal elements, making it intractable
to analytically invert the matrix. The presence of these off-diagonal
elements indicate the information coupling (IC) among different location
states. One of the most significant IC phenomena in LocTrack systems
is the error propagation (EP) in the temporal dimension, where the
estimation error at one time instance can affect subsequent estimations.
Understanding how error propagates across the process helps in designing
more effective LocTrack algorithms to mitigate the error accumulation,
and allows for better allocation of system resources to areas which
have the most impact on the overall error. In existing literature,
IC has been discussed under different settings. For instance, \cite{IC-Xiong}
and \cite{IC-sp-tem} examined IC in cooperative wireless sensor networks
where inter-agent communication is enabled for cooperative localization.
However, the IC in their works arised from the agent cooperation,
not from the spatiotemporal correlations as considered in this paper.
Thus, their framework is not applicable to analyzing IC in MRMU LocTrack
systems, especially concerning EP. Authors in \cite{EP-zhou} considered
a single target tracking problem with a temporal correlation model
and focused on EP analysis. However, their framework cannot be utilized
to analyze IC in our context. 

In this paper, we establish a theoretical framework for analyzing
the relationship between localization performance and STP in MRMU
LocTrack systems. We introduce the Efficiency of Correlation (EoC)
to characterize the STP utilization efficiency, modeling the IC phenomenon
and capturing the role of multiple RISs in cooperative information
enhancement. Unlike the Cramér-Rao Bound (CRB), which focuses solely
on localization accuracy, EoC complements CRB by incorporating both
accuracy and system overhead, providing an indirect measure of energy
efficiency. To visualize how EoC characterizes the impact of STP and
multiple RISs, we interpret it through random walk theory as \textquotedblleft correlation
information routing.\textquotedblright{} We then analyze a recursive
MRMU LocTrack system, examining the spatiotemporal joint evolution
of EoC and the EP phenomenon. We derive the convergence conditions
and points of the EFIM, demonstrating the system\textquoteright s
robustness to sudden changes. Finally, asymptotic analysis reveals
the trade-off between localization accuracy and information utilization
efficiency as STPs evolve dynamically. Simulation results verify
the correctness of our theoretical results. Our main contributions
are summarized as follows:
\begin{itemize}
\item We propose a general spatiotemporal Markov random field (ST-MRF) to
capture the dynamic motion of multiple users in both spatial and temporal
domains. The statistical ST-MRF offers inherent flexibility to describe
a wide range of user correlations across different applications.
\item We propose to decompose the EFIM of user location and characterize
the IC phenomenon in STP-assisted MRMU LocTrack systems using a metric
of EoC. We provide a ``correlation information routing'' interpretation
of EoC by introducing the random walk model (RWM) theory, which offers
a new perspective to elucidate the efficiency of STP utilization and
the impact of multiple RISs.
\item We analyze the EP effect in MRMU LocTrack systems considering the
spatiotemporal joint evolution of EoC. We characterize the EP principle,
as well as its convergence behavior over time and its asymptotic behavior
under three extreme cases.
\end{itemize}
The rest of paper is organized as follows. Section \ref{sec:System-Model}
introduces the signal model, spatiotemporal prior model, and the derivation
of Bayesian CRB (BCRB). Section \ref{sec:IC} analyzes the IC phenomenon
based on the EFIM decomposition and provides the graphical interpretation
for EoC. Section \ref{sec:Error-Propagation-Analysis} presents the
EP phenomenon, analyzing its convergence and asymptotic behavior in
LocTrack systems. Finally, extensive simulation results are provided
to validate the correctness of our theoretical analysis in the Section
\ref{sec:Simulation-Results} and conclusions are drawn in Section
\ref{sec:Conclusion}.

\emph{Notation:} Upper case and lower case bold face letters denote
matrices and vectors, respectively. $\mathbf{A}=\mathrm{Diag}\left(\boldsymbol{\alpha}\right)$
stands for the diagonal matrix with vector $\boldsymbol{\alpha}$
on the diagonal and $\mathbf{\boldsymbol{\alpha}}=\mathrm{diag}\left(\mathbf{A}\right)$
means vector $\boldsymbol{\alpha}$ is extracted from the diagonal
elements of matrix $\mathbf{A}$. $\mathbf{A}=\mathrm{BlockDiag}\left[\mathbf{A}_{1},\mathbf{A}_{2}\right]$
stands for a block-diagonal matrix composed of matrix $\mathbf{A}_{1}$
and $\mathbf{A}_{2}$. $\left[\mathbf{A}\right]_{i,j}$ means the
$(i,j)$-th submatrix of $\mathbf{A}$ with $2\times2$ dimension,
while $\left[\boldsymbol{\alpha}\right]_{i}$ means $i$-th element
of vector $\boldsymbol{\alpha}$. $\textrm{Re}\left\{ \cdot\right\} $
and $\textrm{Im}\left\{ \cdot\right\} $ stand for real part and imaginary
part operator. $\otimes$ and $\odot$ stand for Kronecker product
and Element-wise product. $\mathbf{I}_{L}$ means identity matrix
with $L\times L$ dimension. $\mathbf{1}_{L}$ means all one matrix
with $L\times L$ dimension. $\mathbb{E}_{\boldsymbol{a}}\left[\cdot\right]$
denotes the expectation operator w.r.t. the random vector $\boldsymbol{a}$.
$\boldsymbol{x}\sim\mathcal{N}(\boldsymbol{\mu},\boldsymbol{\Sigma})$
means that $\boldsymbol{x}$ follows a Gaussian distribution with
mean $\boldsymbol{\mu}$ and covariance $\boldsymbol{\Sigma}$. $\text{\ensuremath{\mathbf{D}=\mathrm{BlockDiag}[\{\mathbf{D}_{t,k}\}]}}$
stands for $\text{\ensuremath{\mathbf{D}=\mathrm{BlockDiag}[\mathbf{D}_{1,1}\cdots,\mathbf{D}_{1,K},\cdots,\mathbf{\mathbf{D}}_{T,1},\cdots,\mathbf{D}_{T,K}]}}$.
$\mathcal{X}\backslash a$ stands for the set $\mathcal{X}$ excluding
element $a$. $X\rightarrow\infty$ means the non-zero entries of
$\mathbf{X}$ tend to infinity. 

\begin{figure}[t]
\begin{centering}
\includegraphics[width=3.5in]{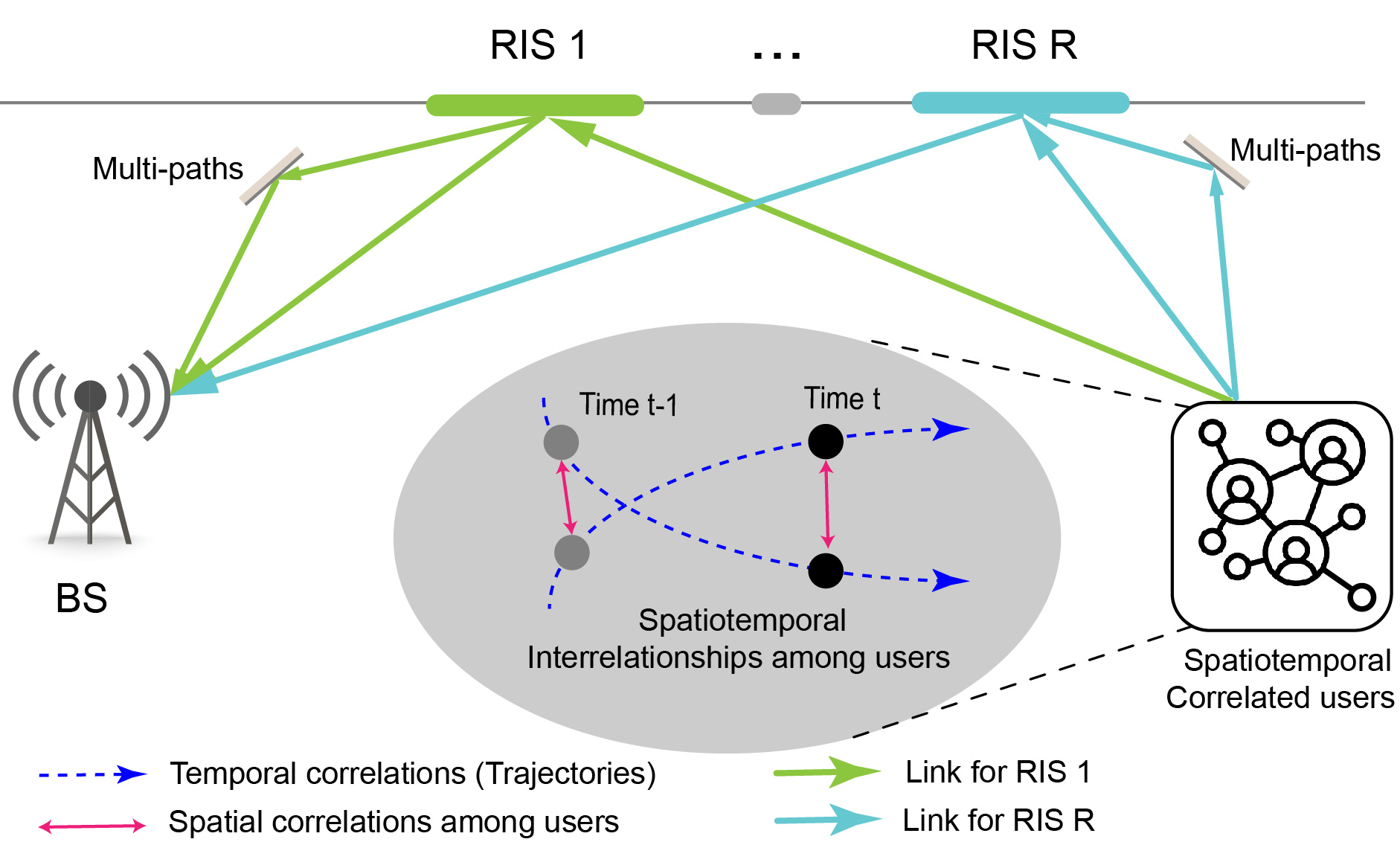}
\par\end{centering}
\caption{Multi-RIS-assisted LocTrack system for spatiotemporal correlated users.
\protect\label{fig:System}}
\end{figure}

\section{System Model\protect\label{sec:System-Model}}

Consider an MRMU LocTrack system, where $K$ single-antenna users
transmit pilot signals to BS reflected via $R$ RISs for LocTrack,
as illustrated in Fig. \ref{fig:System}. The BS is equipped with
$N_{b}$ antennas, and each RIS has $N_{r}$ passive elements. We
consider that the direct links between BS and users are blocked by
obstacles. The positions of BS and the $i$-th RIS are denoted as
$\boldsymbol{b}=\left[b_{x},b_{y}\right]^{T}$ and $\boldsymbol{r}_{i}=\left[r_{x,i},r_{y,i}\right]^{T},\forall i\in\mathcal{R},$
respectively. The position of the $k$-th user at time step $t\in\mathcal{T}$
is denoted as $\boldsymbol{u}_{t,k}=\left[u_{x,t,k},u_{y,t,k}\right]^{T},\forall k\in\mathcal{K}$.
Set $\mathcal{R}$, $\mathcal{K}$ and $\mathcal{T}$ denote the collections
of RIS, user and time horizon indices, respectively. Naturally, the
locations of BS and RISs are fixed and known in advance at the BS.

\subsection{Signal Model}

We establish a general multi-RIS-assisted channel model considering
the multi-path effect as in Fig. \ref{fig:System}. Specifically,
the multi-path channel from the $i$-th RIS to BS can be formulated
by \cite{RIS-channel-MP} 
\begin{equation}
\mathbf{H}_{BR,t,i}=\sqrt{\frac{\kappa_{BR}}{1+\kappa_{BR}}}\mathbf{H}_{BR,t,i}^{(LoS)}+\sqrt{\frac{1}{1+\kappa_{BR}}}\mathbf{H}_{BR,t,i}^{(NLoS)},\label{eq:channel-MP}
\end{equation}
where $\kappa_{BR}$ denotes the multi-path factor between BS and
RISs. $\mathbf{H}_{BR,t,i}^{(NLoS)}$ represents NLoS multi-path components
between BS and the $i$-th RIS, in which the entries can be modeled
as i.i.d. zero-mean Gaussian distribution \cite{RIS-channel-MP}.
The matrix $\mathbf{H}_{BR,t,i}^{(LoS)}$ denotes the LoS path between
BS and the $i$-th RIS, which can be expressed as \cite{RIS-opt-CRB3}
\[
\mathbf{H}_{BR,t,i}^{(LoS)}=\rho_{i}^{BR}\boldsymbol{a}_{B}\left(\theta_{i}^{BR}\right)\boldsymbol{a}_{R}^{H}\left(\varphi_{i}^{BR}\right),
\]
where $\theta_{i}^{BR}$, $\varphi_{i}^{BR}$ and $\rho_{i}^{BR}$
denote AoA, angle-of-departure (AoD) and channel gain from the $i$-th
RIS to BS, respectively. $\boldsymbol{a}_{B}(\cdot)$ and $\boldsymbol{a}_{R}(\cdot)$
denote steering vectors of BS and RIS. Similarly, the multi-path channel
from the $k$-th user to the $i$-th RIS can be formulated as
\[
\boldsymbol{h}_{RU,t,i,k}=\sqrt{\frac{\kappa_{RU}}{1+\kappa_{RU}}}\boldsymbol{h}_{RU,t,i,k}^{(LoS)}+\sqrt{\frac{1}{1+\kappa_{RU}}}\boldsymbol{h}_{RU,t,i,k}^{(NLoS)},
\]
where $\kappa_{RU}$ denotes the multi-path factor between RISs and
users. The NLoS component $\boldsymbol{h}_{RU,t,i,k}^{(NLoS)}$ can
also be modeled as i.i.d. zero-mean Gaussian distribution. LoS component
can be expressed as $\boldsymbol{h}_{RU,t,i,k}^{(LoS)}=\rho_{t,i,k}^{RU}\boldsymbol{a}_{R}\left(\theta_{t,i,k}^{RU}\right)$
with channel gain $\rho_{t,i,k}^{RU}$ and AoA $\theta_{t,i,k}^{RU}$.
Let $\boldsymbol{h}_{t,k}$ denote cascaded uplink channel between
BS and the $k$-th user reflected via $R$ RISs at time $t$, which
can be formulated as
\begin{equation}
\boldsymbol{h}_{t,k}=\sum_{i=1}^{R}\mathbf{H}_{BR,t,i}\mathbf{\boldsymbol{\Omega}}_{t,i}\boldsymbol{h}_{RU,t,i,k},\label{eq:channel=000020model}
\end{equation}
where $\mathbf{\boldsymbol{\Omega}}_{t,i}=\frac{1}{\sqrt{N_{r}}}\textrm{diag}\left(e^{j\omega_{t,i,1}},\ldots,e^{j\omega_{t,i,N_{r}}}\right)\overset{\triangle}{=}\textrm{diag}\left(\boldsymbol{\omega}_{t,i}\right)$
represents the phase control matrix of the $i$-th RIS corresponding
to time $t$. The user location-related parameters $\rho_{t,i,k}^{RU}$
and $\theta_{t,i,k}^{RU}$ can be calculated from the geometrical
relations between RIS and user as follows: 
\begin{equation}
\begin{array}{l}
\theta_{t,i,k}^{RU}=\arccos\left(\left(u_{x,t,k}-r_{x,i}\right)/\left\Vert \boldsymbol{r}_{i}-\boldsymbol{u}_{t,k}\right\Vert \right),\\
\rho_{t,i,k}^{RU}=\left(\left\Vert \boldsymbol{r}_{i}-\boldsymbol{u}_{t,k}\right\Vert \right)^{-\alpha/2},
\end{array}
\end{equation}
where $\alpha$ is path loss exponent.

Denote $P$ pilot symbols transmitted by each user of time $t$ as
$\boldsymbol{x}_{t,k}=[x_{t,k}^{1},\ldots,x_{t,k}^{P}]^{T}$ with
$\mathrm{Tr}\left(\boldsymbol{x}_{t,k}\boldsymbol{x}_{t,k}^{H}\right)=1$.
Assume user location $\boldsymbol{u}_{t,k}$ and NLoS channel are
unchanged during the time interval of $P$ pilots, the received signal
at BS at time step $t$ can be written as
\begin{align}
\mathbf{Y}_{t} & =\sum_{k=1}^{K}\boldsymbol{h}_{t,k}\boldsymbol{x}_{t,k}^{T}+\boldsymbol{N}_{t}=\sum_{k=1}^{K}\bar{\boldsymbol{h}}_{t,k}\boldsymbol{x}_{t,k}^{T}+\tilde{\boldsymbol{N}}_{t},\label{eq:rec_y}
\end{align}
where the entries of noise matrix $\boldsymbol{N}_{t}$ are i.i.d.
distributed, following a Gaussian distribution with zero mean and
variance $\sigma_{t}^{2}$. $\bar{\boldsymbol{h}}_{t,k}$ denotes
the direct link channel, given by 
\[
\bar{\boldsymbol{h}}_{t,k}=\sum_{i=1}^{R}\text{\ensuremath{\rho_{t,i,k}}}\boldsymbol{a}_{B}\left(\theta_{i}^{BR}\right)\boldsymbol{a}_{R}^{H}\left(\varphi_{i}^{BR}\right)\mathbf{\boldsymbol{\Omega}}_{t,i}\boldsymbol{a}_{R}\left(\theta_{t,i,k}^{RU}\right),
\]
with $\text{\ensuremath{\rho_{t,i,k}}}=\sqrt{\frac{\kappa_{BR}\kappa_{RU}}{\left(1+\kappa_{BR}\right)\left(1+\kappa_{RU}\right)}}\rho_{i}^{BR}\rho_{t,i,k}^{RU}$.
$\tilde{\boldsymbol{N}}_{t}$ is multi-path interference-plus-noise
matrix, given by 
\begin{align*}
\tilde{\boldsymbol{N}}_{t} & =\sum_{k=1}^{K}\sum_{i=1}^{R}\left(\xi\mathbf{H}_{BR,t,i}^{(NLoS)}\mathbf{\boldsymbol{\Omega}}_{t,i}\boldsymbol{h}_{RU,t,i,k}^{(NLoS)}\right.\\
 & +\xi_{BR}\mathbf{H}_{BR,t,i}^{(LoS)}\mathbf{\boldsymbol{\Omega}}_{t,i}\boldsymbol{h}_{RU,t,i,k}^{(NLoS)}\\
 & \left.+\xi_{RU}\mathbf{H}_{BR,t,i}^{(NLoS)}\mathbf{\boldsymbol{\Omega}}_{t,i}\boldsymbol{h}_{RU,t,i,k}^{(LoS)}\right)\boldsymbol{x}_{t,k}^{T}+\boldsymbol{N}_{t},
\end{align*}
where $\xi=\sqrt{\frac{1}{\left(1+\kappa_{BR}\right)\left(1+\kappa_{RU}\right)}}$,
$\xi_{BR}=\sqrt{\frac{\kappa_{BR}}{\left(1+\kappa_{BR}\right)\left(1+\kappa_{RU}\right)}}$
and $\xi_{RU}=\sqrt{\frac{\kappa_{RU}}{\left(1+\kappa_{BR}\right)\left(1+\kappa_{RU}\right)}}$.
Note that entries of $\tilde{\boldsymbol{N}}_{t}$ also follow zero
mean complex Gaussian distribution due to the Rician channel assumption.
To extract the observation signals corresponding to each user, we
assume the pilot sequences of different users are orthogonal, i.e.,
$\boldsymbol{x}_{t,k}^{H}\boldsymbol{x}_{t,k'}=0,k\neq k',\forall k,k'\in\mathcal{K}$.
When the pilot sequence length exceeds the number of users involved
in LocTrack, i.e., $P\geq K$, such orthogonal pilot design becomes
straightforward. Therefore, the measurement corresponding to user
$k$ is given by
\begin{equation}
\boldsymbol{y}_{t,k}=\bar{\boldsymbol{h}}_{t,k}+\bar{\boldsymbol{n}}_{t,k}.\label{eq:y_k}
\end{equation}
The goal of MRMU LocTrack is to joint estimate $\{\boldsymbol{u}_{t,k}\}$
from $\{\boldsymbol{y}_{t,k}\}$ exploiting the spatiotemporal correlations
among users' positions.

To assess the impact of STP on LocTrack performance, we first consider
a batch system, which performs smoothing by utilizing both past and
future time steps for position estimation. We analyze this with EFIM
decomposition for EoC derivation and its graphical interpretation
in Section \ref{sec:IC}. Then, we focus on a recursive system, which
performs filtering by relying solely on past time steps for position
estimation. We analyze the recursive representation of EFIM and the
EP phenomenon in Section \ref{sec:Error-Propagation-Analysis}.

\subsection{Statistical Spatiotemporal Correlation Model}

In various sensing applications, multi-user LocTrack performance can
be improved by exploiting the correlations among users' positions
across spatial and temporal domain. We propose to adopt the MRF \cite{Tem-Gau-MRF}
model to characterize the spatiotemporal correlations among users'
positions $\boldsymbol{u}=[\boldsymbol{u}_{1}^{T},\cdots,\boldsymbol{u}_{T}^{T}]^{T}$,
where $\boldsymbol{u}_{t}=[\boldsymbol{u}_{t,1}^{T},\cdots,\boldsymbol{u}_{t,K}^{T}]^{T}$.
The MRF is a factored probability function specified by an undirected
graph $(V,E),$ where $V$ stands for the variable nodes and $E$
specifies the correlation network. If $(i,j)\in E$, which means there
is direct connection between node $i$ and node $j$, and node $i,j$
are correlated with each other. Here we assume the graph is undirected,
which means if $(i,j)\in E$, we have $(j,i)\in E$. The joint probability
over the random variables in MRF can be factorized as the product
of local potential functions $\phi$ at each node and interaction
potentials $\psi$ defined on neighborhood cliques. A widely adopted
MRF model is the pairwise MRF \cite{Tem-Gau-MRF}, where the cliques
are restricted to pairs of nodes. We adopt the pairwise MRF to model
the STP of $\boldsymbol{u}$, called spatiotemporal-MRF (ST-MRF),
given by 
\begin{align}
p\left(\boldsymbol{u}\right)= & p(\boldsymbol{u}_{1})\prod_{t=2}^{T}p(\boldsymbol{u}_{t}|\boldsymbol{u}_{t-1})\nonumber \\
= & \prod_{k=1}^{K}\phi(\boldsymbol{u}_{1,k})\prod_{(i,j)\in E}\varphi\left(\boldsymbol{u}_{1,i},\boldsymbol{u}_{1,j}\right)\nonumber \\
\times & \prod_{t=2}^{T}\prod_{k=1}^{K}p\left(\boldsymbol{u}_{t,k}\mid\boldsymbol{u}_{t-1,k}\right)\prod_{(i,j)\in E}\varphi\left(\boldsymbol{u}_{t,i},\boldsymbol{u}_{t,j}\right).\label{eq:prior_dis}
\end{align}

In \eqref{eq:prior_dis}, the temporal correlations are captured by
the transition probability $p\left(\boldsymbol{u}_{t,k}\mid\boldsymbol{u}_{t-1,k}\right)$,
which is designed to follow a Gaussian distribution \cite{tRIS-tr1,Tem-Gau-MRF,Tem-Gau-MRF2,Tem-Gau-MRF3},
i.e.,

\begin{equation}
p(\boldsymbol{u}_{t,k}\mid\boldsymbol{u}_{t-1,k})=\mathcal{N}(\boldsymbol{u}_{t-1,k},\text{\ensuremath{\boldsymbol{Q}}}_{t,k}),\label{eq:Temproal_corr}
\end{equation}
where $\text{\ensuremath{\boldsymbol{Q}}}_{t,k}$ denotes the covariance
matrix of uncertainty in the position update at time $t$. The local
potential functions \textbf{$\phi(\boldsymbol{u}_{1,k})$} at the
first time step are usually set to be Gaussian \cite{Tem-first-MRF1,Tem-first-MRF2}.
The spatial correlations are modeled by pairwise potential functions
$\varphi\left(\boldsymbol{u}_{t,i},\boldsymbol{u}_{t,j}\right)$,
which are typically designed as the function of inter-user distances
\cite{Toy1-square,Toy2-square,Toy1-norm,Toy2-norm}, i.e., $\varphi\left(\boldsymbol{u}_{t,i},\boldsymbol{u}_{t,j}\right)=p\left(d_{ij}^{t}\right),$
with $d_{ij}^{t}=\left\Vert \boldsymbol{u}_{t,i}-\boldsymbol{u}_{t,j}\right\Vert $.

\begin{figure}[t]
\begin{centering}
\includegraphics[width=3in]{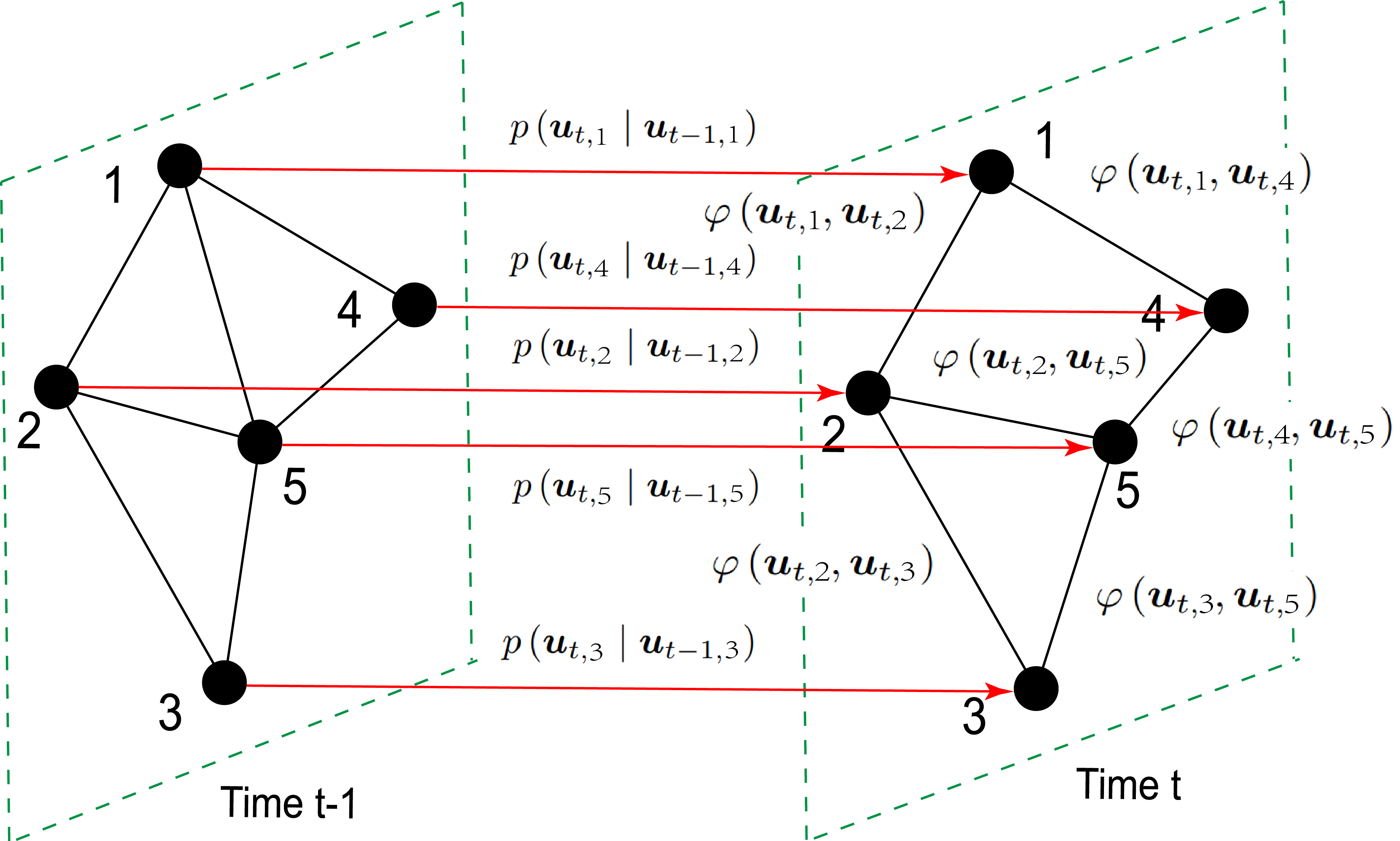}
\par\end{centering}
\caption{An illustration of ST-MRF priors with $K=5$ users. \protect\label{fig:MRF}}
\end{figure}

As shown in Fig. \ref{fig:MRF}, we provide a toy example of ST-MRF
priors in a LocTrack system for $K=5$ users. For example, position
of user 1 exhibits correlations not only with its spatially neighboring
users 2, 4 and 5, but also with its own state from last time step
$t-1$. The position of user 5 demonstrates varying spatial correlations
in different time steps. For example, it is correlated with users
2, 3, and 4 in time step $t-1$, and with users 1, 2, 3, and 4 in
time step $t$. Therefore, the proposed statistical ST-MRF exhibits
flexibility in accommodating a wide range of spatial and temporal
correlations. In the following, we provide two typical probability
models for the pairwise potential function $\varphi$ in \eqref{eq:prior_dis}
that have been widely adopted in localization works \cite{Toy1-square,Toy2-square,Toy1-norm,Toy2-norm}
in different scenarios.
\begin{itemize}
\item \emph{Pedestrian surveillance:} For the pedestrian surveillance problem
\cite{Toy1-square,Toy2-square}, where several groups of pedestrians
or sportsmen walk together in a monitored area, positions of each
group exhibits significant inter-correlation. In this case, the potential
functions $\varphi\left(\boldsymbol{u}_{t,i},\boldsymbol{u}_{t,j}\right)$
in \eqref{eq:prior_dis} is given by the square of $l_{2}$-norm priors
in a exponential function, i.e.,
\begin{equation}
\varphi\left(\boldsymbol{u}_{t,i},\boldsymbol{u}_{t,j}\right)\propto\exp\left(-\frac{1}{2}\frac{\left(d_{ij}^{t}\right)^{2}}{2\sigma_{t,ij}^{2}}\right),\label{eq:Pedestrian}
\end{equation}
where $\sigma_{t,ij}^{2}$ is the variance of inter-user distance.
Note that the factor $\frac{1}{2}$ in the exponential term is used
to eliminate duplicate multiplication for edge $(i,j)\in E$ in \eqref{eq:prior_dis}.
\item \emph{Communities with few distant users:} For most of multi-user
LocTrack problem, the interaction potential functions can be modeled
as a quadratic prior as in \eqref{eq:Pedestrian}. However, the $l_{2}$-norm
priors can limit the penalty computed for users who are far from the
community members, particularly in large sensor networks \cite{Toy1-norm,Toy2-norm}.
To improve the system robustness, $l_{1}$-norm priors can be exploited,
i.e., 
\begin{equation}
\varphi\left(\boldsymbol{u}_{t,i},\boldsymbol{u}_{t,j}\right)\propto\exp\left(-\frac{1}{2}\frac{d_{ij}^{t}}{2\sigma_{t,ij}^{2}}\right).\label{eq:Pedestrian-1}
\end{equation}
\end{itemize}

\subsection{Bayesian Cramer-Rao Bound}

Based on the received signal model in \eqref{eq:y_k} and the STP
in \eqref{eq:prior_dis}, we can obtain the position error bound of
each user at each time step in terms of Bayesian CRB (BCRB), which
is a theoretical lower bound of position mean square error (MSE).
Denote the unknown parameters in a multi-RIS assisted multi-user LocTrack
system as $\boldsymbol{\eta}=[\boldsymbol{u}^{T},\boldsymbol{\xi}^{T}]^{T}$,
where $\boldsymbol{u}$ denotes the collection of users' positions,
$\boldsymbol{\xi}$ denotes the nuisance parameters associated with
RIS-reflected channels, observation noise, etc. Then the lower bound
of position MSE for the unbiased estimator $\hat{\boldsymbol{u}}$
of $\boldsymbol{u}$ is given by 
\begin{equation}
\mathbb{E}_{\boldsymbol{u}}\left[\left(\hat{\boldsymbol{u}}-\boldsymbol{u}\right)\left(\hat{\boldsymbol{u}}-\boldsymbol{u}\right)^{T}\right]\succeq\left[\mathbf{J}_{e}(\boldsymbol{u})\right]^{-1},\label{eq:CRB}
\end{equation}
where $\mathbf{J}_{e}$ is the EFIM of $\boldsymbol{u}$ with respect
to $\boldsymbol{\eta}$. Then the BCRB of $K$ users' positions of
all time steps is defined as
\begin{align}
B & \triangleq\mathrm{Tr}\left(\left[\mathbf{J}_{e}(\boldsymbol{u})\right]^{-1}\right),\label{eq:FIM_def}
\end{align}
To characterize the LocTrack performance of each user, we define the
BCRB of user $k$ at time $t$ as
\begin{equation}
B_{t,k}\triangleq\mathrm{Tr}\left(\left[\mathbf{J}_{e,t,k}(\boldsymbol{u}_{t,k})\right]^{-1}\right),\label{eq:BCRB_tk}
\end{equation}
where $\mathbf{J}_{e,t,k}$ is the EFIM of $\boldsymbol{u}_{t,k}$
with respect to $\boldsymbol{\eta}$. $\mathbf{J}_{e,t,k}$ contains
all necessary information from measurements and STP for $\boldsymbol{u}_{t,k}$,
and can be obtained from $\mathbf{J}_{e}$ through $\mathbf{J}_{e,t,k}^{-1}=\left[\mathbf{J}_{e}^{-1}\right]_{(t-1)K+k,(t-1)K+k}$.
In the following, we give the detailed expression of $\mathbf{J}_{e}$.
\begin{lem}
[EFIM of Users' Positions\label{lem:EFIM}] The EFIM $\mathbf{J}_{e}\in\mathbb{R}^{2TK\times2TK}$
in \eqref{eq:FIM_def} includes complete positional information of
$K$ users for $T$ time steps which can be equivalently expressed
as
\begin{equation}
\mathbf{J}_{e}(\boldsymbol{u})=\boldsymbol{\Lambda}_{D}+\mathbf{\boldsymbol{\Lambda}}_{P},\label{eq:FIM-1}
\end{equation}
where $\boldsymbol{\Lambda}_{D}$ is a block-diagonal EFIM with $2\times2$
sized blocks derived from the observations $\boldsymbol{y}=[\boldsymbol{y}_{1,1},\boldsymbol{y}_{1,2},\ldots,\boldsymbol{y}_{T,K}],$
and $\mathbf{\boldsymbol{\Lambda}}_{P}$ is a non-block-diagonal FIM
derived from the ST-MRF priors $p\left(\boldsymbol{u}\right)$. $\mathbf{\boldsymbol{\Lambda}}_{P}$
can be further divided into two submatrices, i.e.,
\[
\mathbf{\boldsymbol{\Lambda}}_{P}=\mathbf{\boldsymbol{\Lambda}}_{PS}+\mathbf{\boldsymbol{\Lambda}}_{PT},
\]
where $\mathbf{\boldsymbol{\Lambda}}_{PS}$ denotes the FIM from the
spatial priors and $\mathbf{\boldsymbol{\Lambda}}_{PT}$ denotes the
FIM from the temporal priors.

In particular, $\boldsymbol{\Lambda}_{D}$ can be calculated by
\begin{align}
\boldsymbol{\Lambda}_{D} & =\mathbf{T}_{\boldsymbol{u}}\left(\boldsymbol{\Lambda}_{\boldsymbol{\theta}}-\boldsymbol{\Lambda}_{\boldsymbol{\xi}\boldsymbol{\theta}}^{T}\boldsymbol{\Lambda}_{\boldsymbol{\xi}}^{-1}\boldsymbol{\Lambda}_{\boldsymbol{\xi}\boldsymbol{\theta}}\right)\mathbf{T}_{\boldsymbol{u}}^{T},\label{eq:Lambda_D}
\end{align}
where $\mathbf{T}_{\boldsymbol{u}}$, $\boldsymbol{\Lambda}_{\boldsymbol{\theta}}$,
$\boldsymbol{\Lambda_{\boldsymbol{\xi}\boldsymbol{\theta}}}$ and
$\boldsymbol{\Lambda}_{\boldsymbol{\xi}}$ are given in Appendix \ref{subsec:Proof=000020of=000020EFIM}.
For a general STP in \eqref{eq:prior_dis}, $\mathbf{\boldsymbol{\Lambda}}_{PS}$
and $\mathbf{\boldsymbol{\Lambda}}_{PT}$ are given in \eqref{eq:Gamma_PS}
and \eqref{eq:Gamma_PT}. For $\forall i\in\left\{ 1,\ldots,TK\right\} $
and $\mathcal{I}_{i}=\left\{ 1,\ldots,TK\right\} \setminus i$, we
have:

(1) $\mathbf{\boldsymbol{\Lambda}}_{PS}$ is a symmetric and block-diagonal
FIM with $K\times K$ sized blocks, satisfying 
\begin{align*}
\left[\mathbf{\boldsymbol{\Lambda}}_{PS}\right]_{i,i} & =-\sum_{j\in\mathcal{I}_{i}}\left[\mathbf{\boldsymbol{\Lambda}}_{PS}\right]_{i,j}=-\sum_{j\in\{(i,j)\in E\}}\left[\mathbf{\boldsymbol{\Lambda}}_{PS}\right]_{i,j}.
\end{align*}

(2) $\mathbf{\boldsymbol{\Lambda}}_{PT}$ is a symmetric and tridiagonal
block FIM with $K\times K$ sized blocks, satisfying
\begin{align}
\left[\mathbf{\boldsymbol{\Lambda}}_{PT}\right]_{i,i} & =-\sum_{j\in\mathcal{I}_{i}}\left[\mathbf{\boldsymbol{\Lambda}}_{PT}\right]_{i,j}\nonumber \\
 & =-\left[\mathbf{\boldsymbol{\Lambda}}_{PT}\right]_{i,i+K}-\left[\mathbf{\boldsymbol{\Lambda}}_{PT}\right]_{i,i-K},\label{eq:Sum_pro_T}
\end{align}
with $\left[\mathbf{\boldsymbol{\Lambda}}_{PT}\right]_{i,j}=\boldsymbol{0},$
if $j<0$ or $j>K$.
\end{lem}
\begin{IEEEproof}
See Appendix \ref{subsec:Proof=000020of=000020EFIM}.
\end{IEEEproof}
As shown in \eqref{eq:FIM_def}, the lower bound of position MSE requires
the inverse of the EFIM $\mathbf{J}_{e}(\boldsymbol{u})$, which can
be decomposed into the sum of two FIM matrices. $\boldsymbol{\Lambda}_{D}$
represents positional information extracted from observations $\{\boldsymbol{y}_{k,t}\}$,
embedded within the aggregated reflected signals from multiple RISs.
On one hand, leveraging the spatial degrees of freedom provided by
multiple RISs, the aggregation of RIS-reflected signals at the BS
enhances positional information. On the other hand, the passive beamforming
capability of RISs further improves the quality of positional information
within each reflected signal. Therefore, in LocTrack systems, the
deployment of multiple RISs can effectively enhance information utilization
efficiency. Since the BS can observe each user independently, the
off-diagonal elements in $\boldsymbol{\Lambda}_{D}$ are zero. $\boldsymbol{\Lambda}_{P}$
stands for positional information extracted from the STP. Due to spatial
and temporal correlations in the location priors, $\boldsymbol{\Lambda}_{P}$
does not exhibit a block-diagonal structure with respect to each position
state $\boldsymbol{u}_{t,k},\forall t,k$. The off-diagonal elements
in $\boldsymbol{\Lambda}_{P}$ introduce IC, potentially affecting
the efficiency of correlation exploitation. In Section \ref{sec:IC},
we quantify the effectiveness of STP utilization in MRMU LocTrack
systems using the EoC metric, which mathematically models the IC effect.

\section{Information Coupling Analysis and EoC Interpretations\protect\label{sec:IC}}

Due to the non-diagonal structure of EFIM, there exists IC between
different location states. To analyze this IC phenomenon, we present
a decomposition of EFIM inspired by previous work \cite{IC-Xiong},
through which, we introduce an important metric, i.e., EoC, to quantify
the effectiveness of leveraging prior information. Following the similar
approach to \cite{IC-Xiong}, we propose a graphic interpretation
of EoC based on the theory of random walks.

\subsection{EFIM Decomposition for EoC Derivation}

Based on Lemma \ref{lem:EFIM}, the EFIM in \eqref{eq:FIM-1} can
be rewritten as 
\begin{align}
\mathbf{J}_{e}(\boldsymbol{u}) & =\boldsymbol{\Lambda}_{D}+\mathbf{\boldsymbol{\Lambda}}_{PS}^{D}+\mathbf{\boldsymbol{\Lambda}}_{PT}^{D}-\mathbf{A}\overset{\triangle}{=}\mathbf{D}-\mathbf{A}\nonumber \\
 & =\mathbf{D}\left(\mathbf{I}-\mathbf{D}^{-1}\mathbf{A}\right).\label{eq:difference}
\end{align}
$\mathbf{D}=\boldsymbol{\Lambda}_{D}+\mathbf{\boldsymbol{\Lambda}}_{PS}^{D}+\mathbf{\boldsymbol{\Lambda}}_{PT}^{D}$
is a block-diagonal matrix, where each block has dimensions of $2\times2$,
corresponding to each position state $\boldsymbol{u}_{t,k}$, $\forall t,k$,
i.e., $\mathbf{D}=\mathrm{BlockDiag}\left[\{\mathbf{D}_{t,k}\}\right],$
$\boldsymbol{\Lambda}_{D}=\mathrm{BlockDiag}[\{\boldsymbol{\Lambda}_{D,t,k}\}]$,
$\mathbf{\boldsymbol{\Lambda}}_{PS}^{D}=\mathrm{BlockDiag}[\{\mathbf{\boldsymbol{\Lambda}}_{PS,t,k}^{D}\}]$
and $\mathbf{\boldsymbol{\Lambda}}_{PT}^{D}=\mathrm{BlockDiag}[\{\mathbf{\boldsymbol{\Lambda}}_{PT,t,k}^{D}\}]$,
respectively. $\mathbf{\boldsymbol{\Lambda}}_{PS}^{D}$ and $\mathbf{\boldsymbol{\Lambda}}_{PT}^{D}$
correspond to matrices that extract diagonal blocks, each of size
$2\times2$, from matrices $\mathbf{\boldsymbol{\Lambda}}_{PS}$ and
$\mathbf{\boldsymbol{\Lambda}}_{PT}$, respectively. Matrix $\mathbf{A}$
corresponds to the off-diagonal blocks of $\mathbf{\boldsymbol{\Lambda}}_{PS}$
and $\mathbf{\boldsymbol{\Lambda}}_{PT}$. The expressions of $\boldsymbol{\Lambda}_{D,t,k},$$\mathbf{\boldsymbol{\Lambda}}_{PS,t,k}^{D}$,
$\mathbf{\boldsymbol{\Lambda}}_{PT,t,k}^{D}$ and $\mathbf{A}$ can
be found in the Appendix \ref{subsec:Proof=000020of=000020EFIM} straightforwardly. 

Intuitively, in the ST-MRF model, if we view the neighboring states
that are connected to the current state $\boldsymbol{u}_{t,k}$ as
weak anchors, these weak anchors along with the multiple-RISs provide
positional information for $\boldsymbol{u}_{t,k}$. However, due to
the positional uncertainties of neighboring states, the information
provided by weak anchors becomes degraded. This is the IC phenomenon.
We may have
\begin{equation}
\mathbf{J}_{e,t,k}(\boldsymbol{u}_{t,k})=\mathbf{D}_{t,k}\mathbf{E}_{t,k}\preceq\mathbf{D}_{t,k}.\label{eq:aa}
\end{equation}
$\mathbf{D}_{t,k}$ can be interpreted as the nominal position information
(NPI) of $\boldsymbol{u}_{t,k}$ when IC is ignored and all neighboring
states are considered as anchors with perfectly known positions. The
NPI $\mathbf{D}_{t,k}$ contains three parts of information: information
from measurements at BS originated from the aggregation of multiple
RIS-reflected signals, i.e., $\boldsymbol{\Lambda}_{D,t,k}$, information
from spatial domain neighbors, i.e., $\mathbf{\boldsymbol{\Lambda}}_{PS,t,k}^{D}$,
and information from temporal domain neighbors, i.e., $\mathbf{\boldsymbol{\Lambda}}_{PT,t,k}^{D}$.
$\mathbf{E}_{t,k}$ characterizes how efficiently position state $\boldsymbol{u}_{t,k}$
utilizes positional information from neighboring states. From \eqref{eq:difference},
$\mathbf{E}_{t,k}$ depends on $\mathbf{A}$. In the following, we
introduce a lemma inspired by Taylor's expansion to give a detailed
expression for $\mathbf{E}_{t,k}$.
\begin{lem}
[\label{lem:Dec_EFIM}Decomposition of EFIM ] If the inverse of
$\mathbf{J}_{e}$ exists, the EFIM $\mathbf{J}_{e,t,k}$ for $\boldsymbol{u}_{t,k}$
can be decomposed as 
\begin{equation}
\mathbf{J}_{e,t,k}(\boldsymbol{u}_{t,k})=\mathbf{D}_{t,k}\left(\mathbf{I}+\boldsymbol{\Delta}_{t,k}\right)^{-1},\forall k\in\mathcal{K},t\in\mathcal{T},\label{eq:dec_EFIM}
\end{equation}
 where $\boldsymbol{\Delta}_{t,k}=\sum_{n=1}^{\infty}\left[\mathbf{P}^{n}\right]_{\gamma,\gamma}\succeq\mathbf{0}$
with transition matrix $\mathbf{P}$ given by
\begin{equation}
\mathbf{P}=\left[\begin{array}{cc}
\mathbf{Q} & \mathbf{R}\\
\mathbf{0} & \mathbf{I}_{2}
\end{array}\right].\label{eq:Absorb}
\end{equation}
$\gamma=(t-1)K+k$, $\mathbf{Q}$ and $\mathbf{R}$ are given by $\text{\ensuremath{\mathbf{Q}}}=\mathbf{D}^{-1}\mathbf{A}\in\mathbb{R}^{2TK\times2TK}$
and $\mathbf{R}=\mathbf{D}^{-1}\left[\boldsymbol{\Lambda}_{D,1,1}^{T},\boldsymbol{\Lambda}_{D,1,2}^{T}\cdots,\boldsymbol{\Lambda}_{D,T,K}^{T}\right]^{T}\in\mathbb{R}^{2TK\times2}$,
respectively.
\end{lem}
\begin{IEEEproof}
See Appendix \ref{subsec:Pf-Lemma-Dec}.
\end{IEEEproof}
From the decomposition of EFIM in \eqref{eq:dec_EFIM}, the EoC for
estimating $\text{\ensuremath{\boldsymbol{u}}}_{t,k}$ can be determined
by
\begin{equation}
\mathbf{E}_{t,k}=\left(\mathbf{I}+\boldsymbol{\Delta}_{t,k}\right)^{-1},\label{eq:=000020def_E_t_k}
\end{equation}
which characterizes the efficiency of correlation prior exploitation.
Since $\boldsymbol{\Delta}_{t,k}$ is non-negative, we have 
\[
\mathbf{0}\prec\mathbf{E}_{t,k}\preceq\mathbf{I},
\]
which means that the IC can decrease the information gained at state
$\boldsymbol{u}_{t,k}$, thereby impairing localization performance
and inefficiently consuming system resources in acquiring redundant
information.  In particular, when the overall information derived
from multiple RIS-reflected measurements and neighboring states is
perfectly utilized at BS for localization without positional uncertainties,
it results in $\mathbf{A}=\mathbf{0}$, leading to $\boldsymbol{\Delta}_{t,k}=\mathbf{0}$,
$\mathbf{E}_{t,k}=\mathbf{I}$ and $\mathbf{J}_{e,t,k}=\mathbf{D}_{t,k}$.
This indicates that there is no IC among location states, with each
neighboring state serving as a perfect anchor, thereby achieving full
EoC. Conversely, when the spatial or temporal prior information provided
by neighboring states exhibits high positional uncertainties, we may
have a smaller $\mathbf{E}_{t,k}$, indicating high IC and poor EoC.

\subsection{Graphical Interpretation for EoC}

\begin{figure}[t]
\begin{centering}
\includegraphics[width=3in]{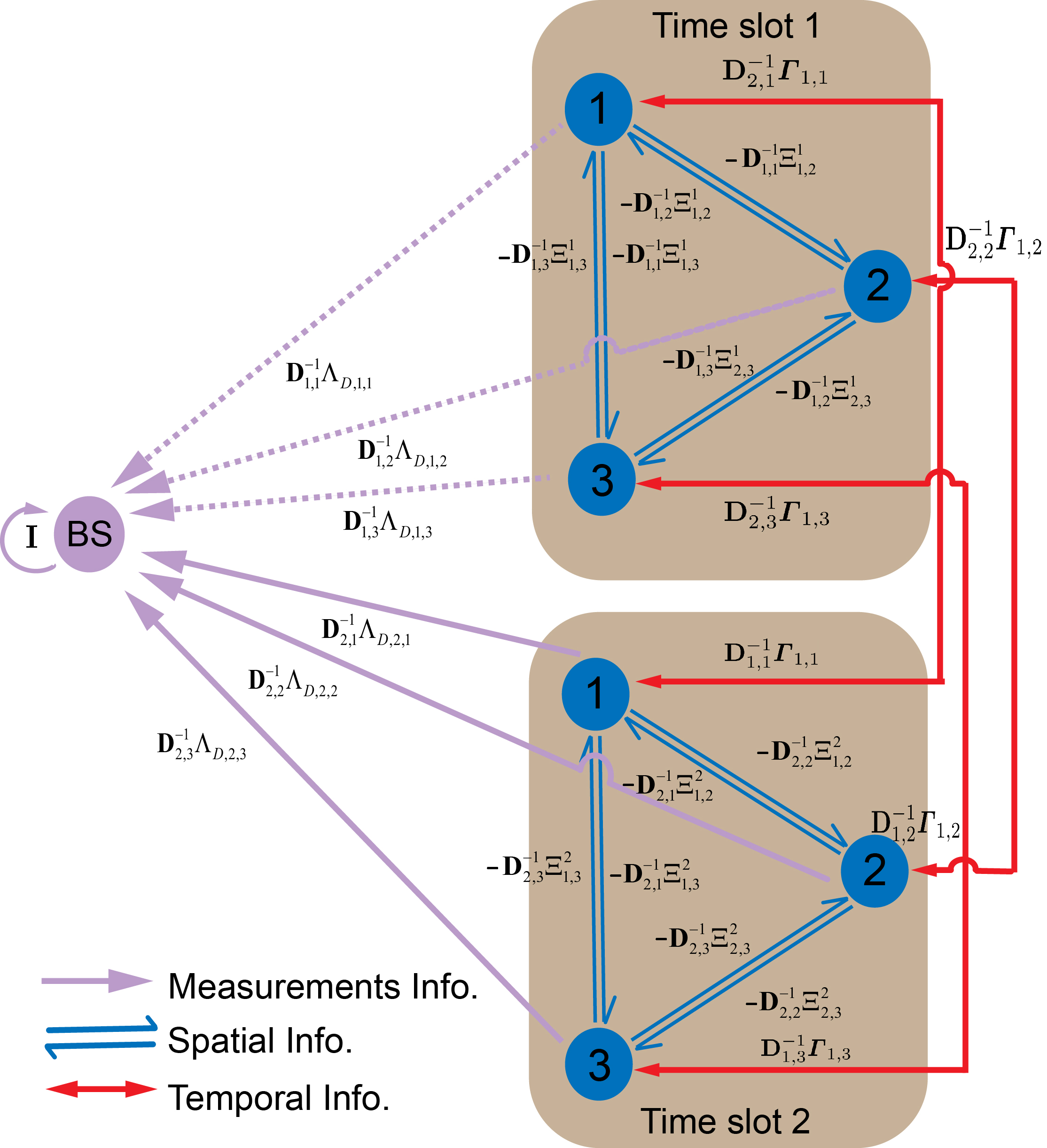}
\par\end{centering}
\caption{Graph interpretation for EoC with $K=3$ users in $T=2$ time steps.
\protect\label{fig:RWM}}
\end{figure}

The EoC $\mathbf{E}_{t,k}$ in \eqref{eq:=000020def_E_t_k} depends
on STP model \eqref{eq:prior_dis}, configurations of multiple RISs
and parameters of the LocTrack systems. To reveal the relationship
between EoC and model/system parameters, we adopt RWM to parse EoC
and provide a graphical interpretation of EoC, depicting the correlation
information routing between different states. In particular, for matrix
$\mathbf{P}$ given in \eqref{eq:Absorb}, it can be observed that
$\mathbf{P}\tilde{\mathbf{I}}=\tilde{\mathbf{I}}$, where $\tilde{\mathbf{I}}=\left[\mathbf{I}_{2},\cdots,\mathbf{I}_{2}\right]^{T}\in\mathbb{R}^{(2TK+2)\times2}.$
While $\mathbf{P}$ does not rigorously satisfy the properties of
a probability measure, by treating $\mathbf{P}$ as a pseudo-transition
probability matrix (PTPM) of a random walk with matrix-valued probabilities
$[\mathbf{P}]_{i,j}$, $1\leq i,j\leq2TK+2$, we can obtain an elegant
interpretation of EoC adopting the classical random walk theory. Denote
$\mathcal{N}=\{(t,k):t\in\mathcal{T};k\in\mathcal{K}\}$ as the collection
of location states at spatiotemporal domain. $\mathcal{B}=\{(t,k):t=T+1;k=1\}$
represents the multi-RIS-enhanced state of the BS, which is primarily
determined by the aggregation of information from multiple RISs. We
have the following proposition.
\begin{prop}
[RWM of EoC \cite{IC-Xiong}\label{prop:RWM-of-EoSCThe}] Considering
a Markov Chain, which consists of a state space $\mathcal{S}=\mathcal{N}\cup\mathcal{B}$
together with a family of random variables $X_{0},X_{1},X_{2},\cdots$
with values in $\mathcal{S}$, the term $\boldsymbol{\Delta}_{t,k}$
can be expressed as
\begin{equation}
\boldsymbol{\Delta}_{t,k}=\sum_{n=1}^{\infty}\mathbb{P}\left(X_{n}=(t,k)|X_{0}=(t,k)\right),\label{eq:Delta_k}
\end{equation}
where $\mathbb{P}\left(X_{n}=(t',k')|X_{0}=(t,k)\right)$ is the $n$-step
PTPM of the Markov chain with the following one-step PTPM:
\begin{align}
 & \mathbb{P}\left(X_{n}=(t',k')|X_{n-1}=(t,k)\right)=\left[\mathbf{P}\right]_{(t-1)*K+k,(t'-1)*K+k'}\nonumber \\
= & \begin{cases}
-\mathbf{D}_{t,k}^{-1}\boldsymbol{\Xi}_{k,k'}^{t} & (t',k')\in\mathcal{N}_{(t,k)}^{S},(t,k)\notin\mathcal{B},\\
\mathbf{D}_{t,k}^{-1}\boldsymbol{\varGamma}_{t',k} & t'=t-1,k'=k,t\in\mathcal{T}\backslash1,\\
\mathbf{D}_{t,k}^{-1}\boldsymbol{\varGamma}_{t'-1,k} & t'=t+1;k'=k,t\in\mathcal{T}\backslash T,\\
\mathbf{D}_{t,k}^{-1}\boldsymbol{\Lambda}_{D,t,k} & (t',k')\in\mathcal{B},(t,k)\notin\mathcal{B},\\
\mathbf{I}_{2} & (t',k')\in\mathcal{B},(t,k)\in\mathcal{B},\\
\mathbf{0} & \mathrm{otherwise},
\end{cases}\label{eq:transition}
\end{align}
where $\mathcal{N}_{(t,k)}^{S}=\{(t',k'):t'=t;k'\neq k,\varphi\left(\boldsymbol{u}_{t,k},\boldsymbol{u}_{t,k'}\right)\neq0\}$
denotes the spatial neighbors of node $(t,k)$. The $n$-step PTPM
can be calculated by 
\begin{align*}
 & \mathbb{P}\left(X_{n}=(t',k')|X_{0}=(t,k)\right)\\
 & =\sum_{(i,j)\in\mathcal{S}}\mathbb{P}\left(X_{n}=(t',k')|X_{n-1}=(i,j)\right)\\
 & \times\mathbb{P}\left(X_{n-1}=(i,j)|X_{0}=(t,k)\right).
\end{align*}
\end{prop}
By the RWM, $\boldsymbol{\Delta}_{t,k}$ can be expressed as 
\begin{equation}
\boldsymbol{\Delta}_{t,k}=\sum_{n=1}^{\infty}[\mathbf{P}^{n}]_{(t-1)*K+k,(t-1)*K+k},\label{eq:=000020RWM_delta}
\end{equation}
which can be interpreted as the sum of $n$-step return probabilities
from state $(t,k$) to itself, where $n=1,2,\ldots,\infty$, $(t,k)\in\mathcal{S}$.
Based on the starting state and the destination state, the transition
probability presents different forms. Specifically, multi-RIS-enhanced
BS state corresponds to the absorbing state of the random walk since
the probability of staying at the state representing the BS is 1.
The location states can transition between their spatial and temporal
neighbors. However, once a transition to the BS state occurs, the
process will remain at the BS state indefinitely. To facilitate understanding,
we here provide a toy example of RWM interpretation of $\boldsymbol{\Delta}_{t,k}$
with $K=3$ users in $T=2$ time steps. In this case, the PTPM matrix
$\mathbf{P}$ takes the form as shown in \eqref{eq:ex_toy-1}, which
can be interpreted by the random walk graph in Fig. \ref{fig:RWM}.
Note that the value beside each edge represents the corresponding
pseudo-transition probability from one state to another. 
\begin{figure*}[t]
\begin{equation}
\mathbf{P}=\left[\begin{array}{c}
\begin{array}{cccccc|c}
\mathbf{0} & -\mathbf{D}_{1,1}^{-1}\boldsymbol{\Xi}_{1,2}^{1} & -\mathbf{D}_{1,1}^{-1}\boldsymbol{\Xi}_{1,3}^{1} & \mathbf{D}_{1,1}^{-1}\boldsymbol{\varGamma}_{1,1} & \mathbf{0} & \mathbf{0} & \mathbf{D}_{1,1}^{-1}\boldsymbol{\Lambda}_{D,1,1}\\
-\mathbf{D}_{1,2}^{-1}\boldsymbol{\Xi}_{1,2}^{1} & \mathbf{0} & -\mathbf{D}_{1,2}^{-1}\boldsymbol{\Xi}_{2,3}^{1} & \mathbf{0} & \mathbf{D}_{1,2}^{-1}\boldsymbol{\varGamma}_{1,2} & \mathbf{0} & \mathbf{D}_{1,2}^{-1}\boldsymbol{\Lambda}_{D,1,2}\\
-\mathbf{D}_{1,3}^{-1}\boldsymbol{\Xi}_{1,3}^{1} & -\mathbf{D}_{1,3}^{-1}\boldsymbol{\Xi}_{2,3}^{1} & \mathbf{0} & \mathbf{0} & \mathbf{0} & \mathbf{D}_{1,3}^{-1}\boldsymbol{\varGamma}_{1,3} & \mathbf{D}_{1,3}^{-1}\boldsymbol{\Lambda}_{D,1,3}\\
\mathbf{D}_{2,1}^{-1}\boldsymbol{\varGamma}_{1,1} & \mathbf{0} & \mathbf{0} & \mathbf{0} & -\mathbf{D}_{2,1}^{-1}\boldsymbol{\Xi}_{1,2}^{2} & -\mathbf{D}_{2,1}^{-1}\boldsymbol{\Xi}_{1,3}^{2} & \mathbf{D}_{2,1}^{-1}\boldsymbol{\Lambda}_{D,2,1}\\
\mathbf{0} & \mathbf{D}_{2,2}^{-1}\boldsymbol{\varGamma}_{1,2} & \mathbf{0} & -\mathbf{D}_{2,2}^{-1}\boldsymbol{\Xi}_{1,2}^{2} & \mathbf{0} & -\mathbf{D}_{2,2}^{-1}\boldsymbol{\Xi}_{2,3}^{2} & \mathbf{D}_{2,2}^{-1}\boldsymbol{\Lambda}_{D,2,2}\\
\mathbf{0} & \mathbf{0} & \mathbf{D}_{2,3}^{-1}\boldsymbol{\varGamma}_{1,3} & -\mathbf{D}_{2,3}^{-1}\boldsymbol{\Xi}_{1,3}^{2} & -\mathbf{D}_{2,3}^{-1}\boldsymbol{\Xi}_{2,3}^{2} & \mathbf{0} & \mathbf{D}_{2,3}^{-1}\boldsymbol{\Lambda}_{D,2,3}\\
\hline \mathbf{0} & \mathbf{0} & \mathbf{0} & \mathbf{0} & \mathbf{0} & \mathbf{0} & \mathbf{I}
\end{array}\end{array}\right],\label{eq:ex_toy-1}
\end{equation}

\rule[0.5ex]{2\columnwidth}{0.5pt}
\end{figure*}

With the above RWM understanding of EoC, based on the random walk
theory in \cite{RWM}, we have the following proposition.
\begin{prop}
[Correlation Information Routing\label{prop:hitting=000020p}]For
the location $\boldsymbol{u}_{t,k}$, its EoC $\mathbf{E}_{t,k}$
can be written as 
\begin{equation}
\mathbf{E}_{t,k}=\left(\mathbf{I}+\boldsymbol{\Delta}_{t,k}\right)^{-1}=\mathbf{I}-\mathbf{F}_{t,k},\label{eq:F1}
\end{equation}
where $\mathbf{F}_{t,k}=\mathbb{P}\left(X_{n}=(t,k),\exists n\geq0|X_{0}=(t,k)\right)$
stands for pseudo-hitting probability \cite{Sec3-hitting} of state
$(t,k)$. 
\end{prop}
If a network only consists of users with unknown locations without
the assistance of multiple RISs and BS, the corresponding RWM is recurrent,
i.e., $\mathbf{F}_{t,k}=\mathbf{I},\forall t,k$. It implies if there
are RISs-BS links that are capable of reflecting, receiving and processing
the location-bearing measurements to effectively extract the location
information, the random walk starting from the state $(t,k$) in $\mathcal{N}$
would always reach the state in $\{(t,k)\cup\mathcal{B}\}$ at least
once. The pseudo-hitting probability $\mathbf{F}_{t,k}$ thus satisfies
\begin{equation}
\mathbf{F}_{t,k}+\mathbf{F}_{t,k\rightarrow\mathcal{B}}=\mathbf{I},\label{eq:F2}
\end{equation}
where $\mathbf{F}_{t,k\rightarrow\mathcal{B}}$ denotes the pseudo-probability
that random walk starting from $(t,k)$ would reach the BS node within
finite time. Combining \eqref{eq:F1} and \eqref{eq:F2}, we have
\begin{equation}
\mathbf{E}_{t,k}=\left(\mathbf{I}+\boldsymbol{\Delta}_{t,k}\right)^{-1}=\mathbf{F}_{t,k\rightarrow\mathcal{B}},\label{eq:F3}
\end{equation}
and 
\begin{equation}
\mathbf{D}_{t,k}=\mathbf{D}_{t,k}\mathbf{F}_{t,k}+\underbrace{\mathbf{D}_{t,k}\mathbf{F}_{t,k\rightarrow\mathcal{B}}}_{\mathbf{J}_{e,t,k}}.\label{eq:F4}
\end{equation}

Therefore, EoC can be interpreted as the efficiency of correlational
position information routing. Specifically, as shown in \eqref{eq:F4},
the NPI $\mathbf{D}_{t,k}$ can be divided into two parts: information
loss $\mathbf{D}_{t,k}\mathbf{F}_{t,k}$ caused by IC and effective
information $\mathbf{J}_{e,t,k}=\mathbf{D}_{t,k}\mathbf{F}_{t,k\rightarrow\mathcal{B}}$
coming from the effective position information routing. In the graph
interpretation, the probability of information loss $\mathbf{F}_{t,k}$
can be viewed as the information routing starting from node $(t,k)$
and returning to itself, which is represented by red and blue lines
in Fig. \ref{fig:RWM}. While the probability of effective information
$\mathbf{F}_{t,k\rightarrow\mathcal{B}}$ can be seen as the information
routing starting from node $(t,k)$ and returning to multi-RIS-enhanced
BS node, which is represented by purple lines. Due to the absence
of location priors at the user nodes, the information routing among
user nodes fails to provide effective location information. In contract,
the BS node, which aggregates signals reflected by multiple virtual
anchors, i.e., RISs, can facilitate the localization process by providing
effective location information.

\subsection{Factors Affecting EoC}

There are primarily two factors that play dominant roles in EoC: STP
priors and measurement model. Consider a simplified scenario with
$K=2$ users over $T=2$ time steps. Assume the spatial correlation
priors are given as $\varphi\left(\boldsymbol{u}_{t,i},\boldsymbol{u}_{t,j}\right)\propto\exp\left(-\frac{1}{2}\frac{\left(d_{ij}^{t}\right)^{2}}{2\sigma_{s}^{2}}\right),$
and the covariance matrix for temporal transitions is given as $\text{\ensuremath{\boldsymbol{Q}}}_{t,k}=\sigma_{t}^{2}\mathbf{I}$
in the ST-MRF $p(\boldsymbol{u})$. From the definition of EoC in
\eqref{eq:=000020def_E_t_k}, EoC $\mathbf{E}_{t,k}$ can be written
as 
\begin{align*}
\mathbf{E}_{t,k} & =\left[\left[\begin{array}{cccc}
\mathbf{I} & -\mathbf{\Theta}_{1,1}^{s} & -\mathbf{\Theta}_{1,1}^{t} & \mathbf{0}\\
-\mathbf{\Theta}_{1,2}^{s} & \mathbf{I} & \mathbf{0} & -\mathbf{\Theta}_{1,2}^{t}\\
-\mathbf{\Theta}_{2,1}^{t} & \mathbf{0} & \mathbf{I} & -\mathbf{\Theta}_{2,1}^{s}\\
\mathbf{0} & -\mathbf{\Theta}_{2,2}^{t} & -\mathbf{\Theta}_{2,2}^{s} & \mathbf{I}
\end{array}\right]_{\gamma,\gamma}^{-1}\right]^{-1},
\end{align*}
with $\gamma=(t-1)K+k$, $\mathbf{D}_{t,k}=\sigma_{n}^{-2}\bar{\boldsymbol{\Lambda}}_{D,t,k}+(\sigma_{s}^{-2}+\sigma_{t}^{-2})\mathbf{I}$
for $t,k\in\{1,2\}$ and $\mathbf{\Theta}_{t,k}^{\mathcal{S}}=\sigma_{\mathcal{S}}^{-2}\mathbf{D}_{t,k}^{-1}$
for set $\mathcal{S}=\{s,t,n\}$, where $\bar{\boldsymbol{\Lambda}}_{D,t,k}$
represents the EFIM that scales $\sigma_{n}^{-2}$ to yield $\boldsymbol{\Lambda}_{D,t,k}$.
Meanwhile, based on \eqref{eq:Absorb}, we can obtain the expression
of PTPM, i.e., 

\[
\mathbf{P}=\left[\begin{array}{c}
\begin{array}{cc|c}
\begin{array}{cc}
\mathbf{0} & \mathbf{\Theta}_{1,1}^{s}\\
\mathbf{\Theta}_{1,2}^{s} & \mathbf{0}
\end{array} & \begin{array}{cc}
\mathbf{\Theta}_{1,1}^{t} & \mathbf{0}\\
\mathbf{0} & \mathbf{\Theta}_{1,2}^{t}
\end{array} & \begin{array}{c}
\mathbf{\Theta}_{1,1}^{n}\boldsymbol{\bar{\Lambda}}_{D,1,1}\\
\mathbf{\Theta}_{2,1}^{n}\boldsymbol{\bar{\Lambda}}_{D,1,2}
\end{array}\\
\begin{array}{cc}
\mathbf{\Theta}_{2,1}^{t} & \mathbf{0}\\
\mathbf{0} & \mathbf{\Theta}_{2,2}^{t}
\end{array} & \begin{array}{cc}
\mathbf{0} & \mathbf{\Theta}_{2,1}^{s}\\
\mathbf{\Theta}_{2,2}^{s} & \mathbf{0}
\end{array} & \begin{array}{c}
\mathbf{\Theta}_{1,1}^{n}\bar{\boldsymbol{\Lambda}}_{D,2,1}\\
\mathbf{\Theta}_{1,1}^{n}\bar{\boldsymbol{\Lambda}}_{D,2,2}
\end{array}\\
\hline \mathbf{0} & \mathbf{0} & \mathbf{I}
\end{array}\end{array}\right].
\]

\begin{figure}[t]
\centering{}\includegraphics[width=3in]{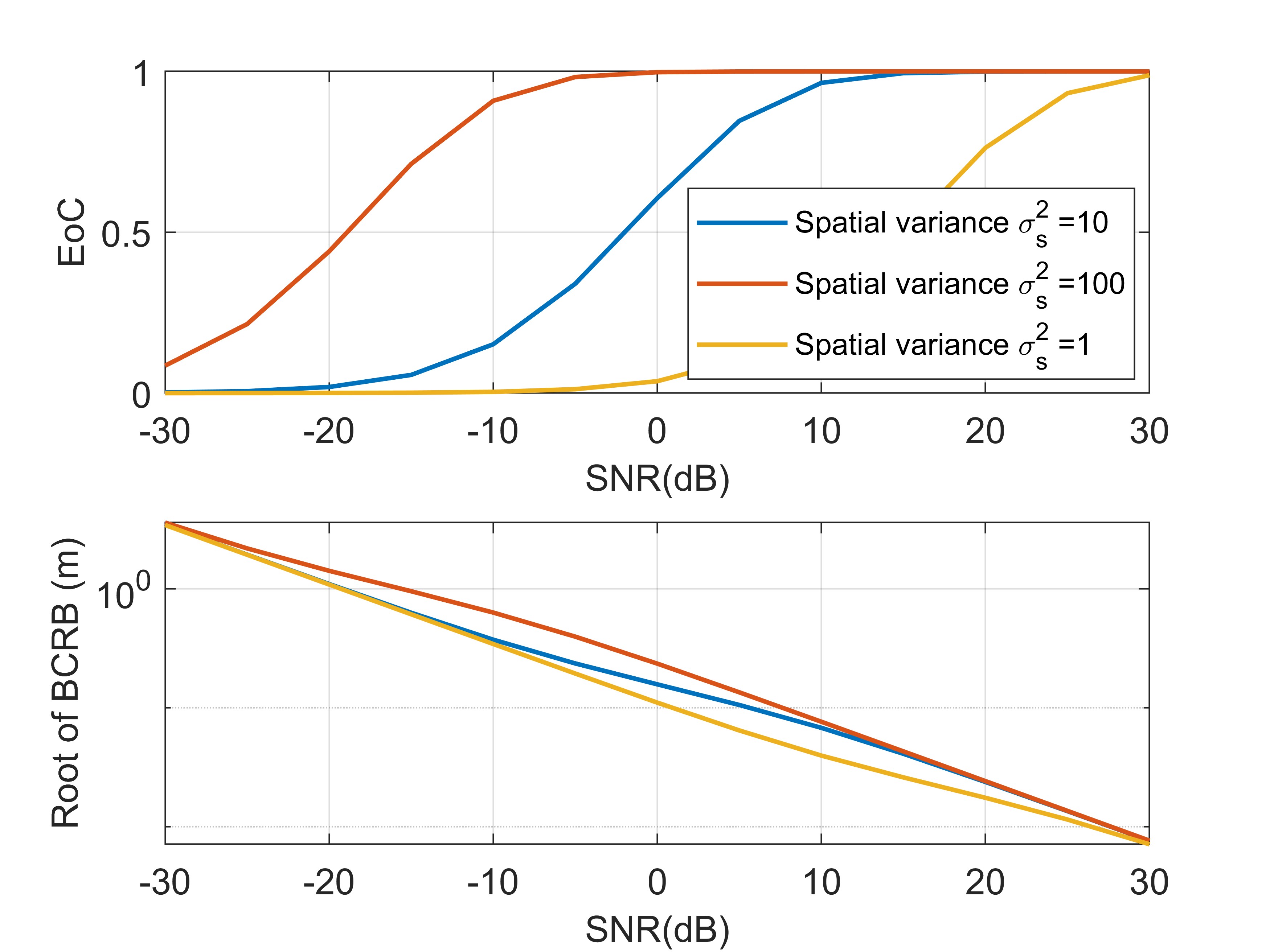}\caption{EoC and BCRB performance versus SNR for different spatial correlation
variances. \protect\label{fig:EoC_RWM_sp-1}}
\end{figure}

\begin{figure}[t]
\centering{}\includegraphics[width=3in]{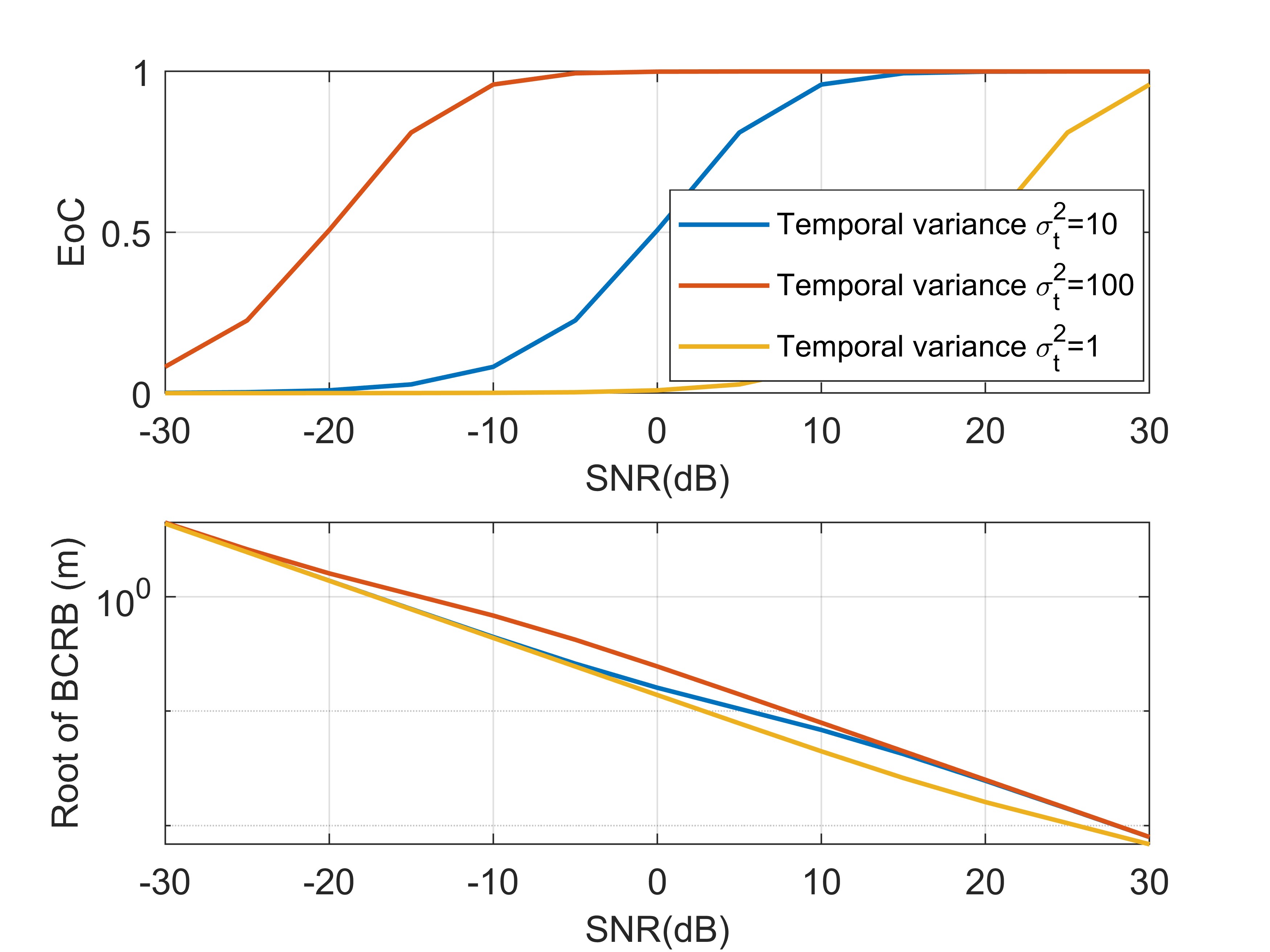}\caption{EoC and BCRB performance versus SNR for different temporal correlation
variances.\protect\label{fig:EoC_RWM_tm-1}}
\end{figure}

\begin{figure}[t]
\includegraphics[width=3in]{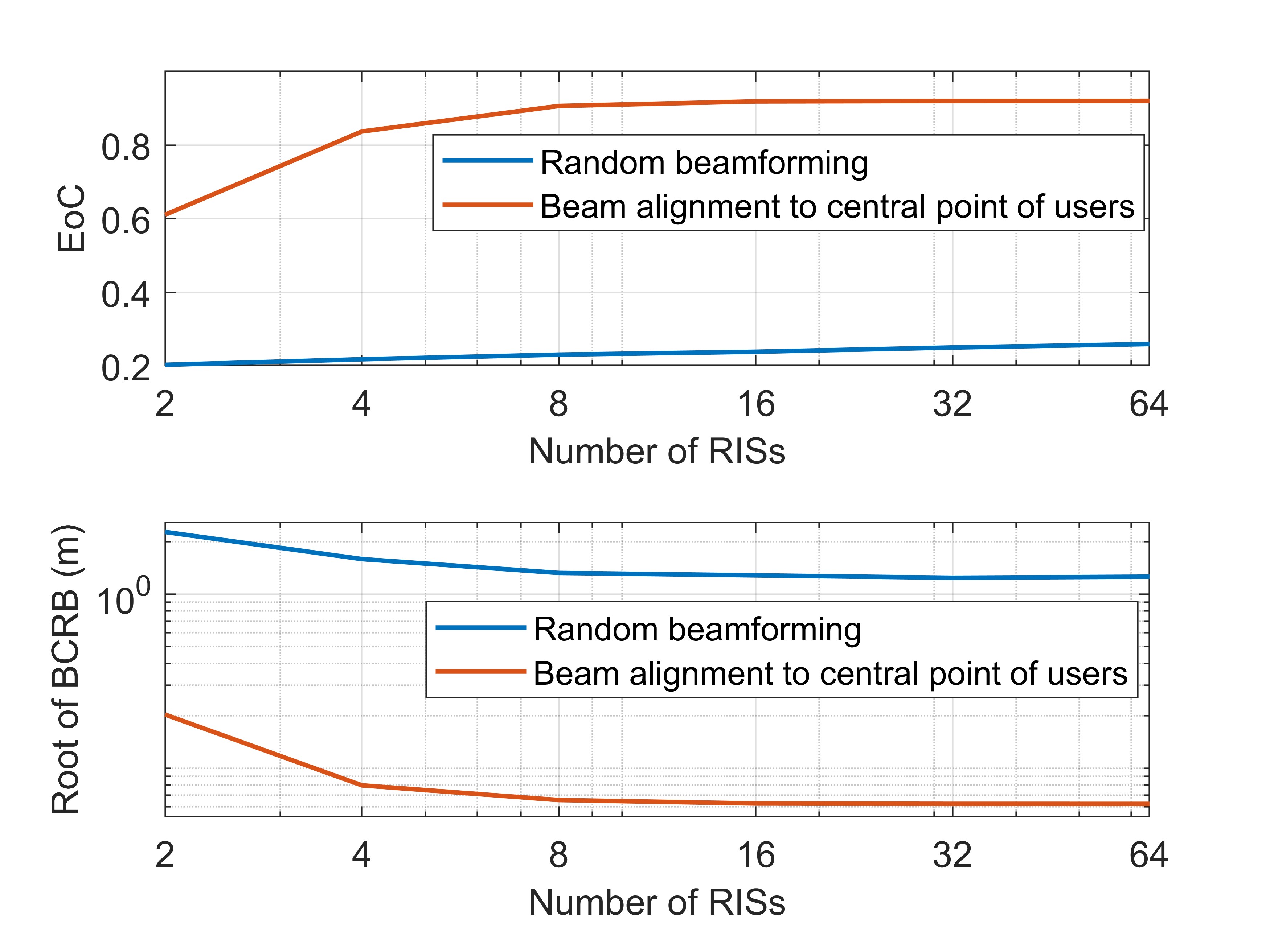}
\centering{}\caption{EoC and BCRB performance versus the number of RISs for different passive
beamforming designs. Blue line stands for random passive beamforming
at multiple RISs. Red line stands for aligning the beams of RISs toward
the users. \protect\label{fig:EoC_RWM_tm-1-1}}
\end{figure}

From the expression of EoC, four primary parameters affect EoC: SNR
of the measurements $\sigma^{-2},$ configurations of multiple RISs
$\bar{\boldsymbol{\Lambda}}_{D,t,k}$, the degree of spatial correlation
$\sigma_{s}^{-2}$ and the degree of temporal correlation $\sigma_{t}^{-2}$.
Given the complexity of the expression, it is hard to directly derive
the relationships between EoC and these influencing factors. In Fig.
\ref{fig:EoC_RWM_sp-1}, Fig. \ref{fig:EoC_RWM_tm-1} and Fig. \ref{fig:EoC_RWM_tm-1-1},
we numerically illustrate their relationships with $\sum_{t,k}\mathrm{Tr}(\mathbf{E}_{t,k})/2TK$,
which is used as an indication of EoC value. For comparison, we also
present the results of the commonly used estimation error metric,
the BCRB, in Fig. \ref{fig:EoC_RWM_sp-1}, Fig. \ref{fig:EoC_RWM_tm-1}
and Fig. \ref{fig:EoC_RWM_tm-1-1}.
\begin{itemize}
\item Impact of SNR: It shows that EoC increases as SNR increases. When
the information obtained from measurements predominates, the impact
of IC is mitigated, thereby enhancing the overall efficiency. From
the perspective of random walk interpretation, $\mathbf{D}_{t,k}^{-1}$
decreases as SNR increases, leading to smaller transition probabilities
between different location states $\mathcal{N}$ and larger transition
probability between location state $\mathcal{N}$ and BS node $\mathcal{B}$.
This suggests a reduction in the positional information flow between
location states, with a corresponding increase in the information
directed toward the BS node. Consequently, this results in the provision
of more effective location information. Meanwhile, an increase in
SNR implies higher quality of information provided by measurements,
thereby leading to a reduction in the BCRB.
\item Configurations of multiple RISs: The impact of RIS is primarily twofold:
the number of RISs $R$ and their passive beamforming matrices $\Omega$.
First, the number of RISs determines the spatial degrees of freedom
for localization. More RISs provide greater spatial freedom, enabling
the extraction of more position-related information from the aggregated
observations at the BS. Second, RISs enhance signal quality at the
BS through passive beamforming, improving the SNR of position-bearing
signals. Together, these factors increase the effectiveness of position
information extraction. From a random walk perspective, as the number
of RISs increases and their beamforming capabilities improve, the
information flow between location states $\mathcal{N}$ decreases,
while the flow towards effective BS nodes increases, thereby enhancing
both information utilization efficiency and localization accuracy.
As illustrated in Fig. \ref{fig:EoC_RWM_tm-1-1}, increasing the number
of RISs, particularly when the total number is small, substantially
improves both localization accuracy and system efficiency. Furthermore,
designing the passive beamforming of RISs by aligning them toward
users significantly improves performance compared to the case without
beamforming design.
\item Impact of correlation priors: It shows that EoC increases as degree
of correlation reduces. Given the quality of measurements, while strong
correlations may provide additional information that results in lower
estimation error for the current location state, such strong correlations
can also lead to an increased reliance on adjacent position states,
whose positions possess positional uncertainties. Consequently, this
results in more pronounced IC, thereby reducing the efficiency of
information utilization. From the perspective of random walk interpretation,
high $\sigma_{s}^{-2}$ (or $\sigma_{t}^{-2}$) will improve the pseudo
probability $\sigma_{s}^{-2}\mathbf{D}_{t,k}^{-1}$ (or $\sigma_{t}^{-2}\mathbf{D}_{t,k}^{-1}$)
and decrease $\mathbf{D}_{t,k}^{-1}\boldsymbol{\Lambda}_{D,t,k}$.
This causes most information being trapped between location states,
hindering effective transmission to the BS node and reducing information
efficiency.
\end{itemize}
Therefore, the impact of different parameters on BCRB and EoC is different,
revealing a trade-off or synergistic relationships between these two
performance metrics. Achieving a balanced consideration of both aspects
in system design is crucial for enhancing the localization accuracy
and optimizing resource utilization efficiency.


\begin{rem}
[How to use EoC for practical design] From the above example, it's
evident that although estimation error metrics, such as BCRB, are
widely used in ISAC as sensing indicators, they primarily focus on
positioning accuracy and neglect the issue of information utilization
efficiency while improving the accuracy. Low EoC indicates that the
performance improvement achieved per unit of resource consumption
is minor, thus operating the system at low EoC is highly uneconomical.
Therefore, a wise transmission scheme for LocTrack system is to jointly
consider the BCRB and EoC simultaneously, instead of minimizing BCRB
as only one criteria. For example, in the MRMU LocTrack systems, we
can design the transmission scheme, such as transmission beamforming
vectors, passive beamforming vectors at multiple RISs, by solving
the following EoC constrained optimization problem:
\begin{align}
 & \min_{\boldsymbol{\omega}}\:BCRB(\boldsymbol{\omega},\boldsymbol{u})\nonumber \\
 & s.t.\quad\mathbf{E}_{t,k}(\boldsymbol{\omega},\boldsymbol{u})\succeq\mathbf{E}_{th},\label{eq:Prob-1}\\
 & \qquad\left|\boldsymbol{\omega}\right|_{i}=1,\nonumber 
\end{align}
where $\boldsymbol{\omega}$ denotes the transmission variables. $\mathbf{E}_{th}$
is the EoC threshold determined by the performance requirements of
the system and the availability of system resources.
\end{rem}

\section{Error Propagation Analysis and Asymptotic Discussion in A Recursive
LocTrack System \protect\label{sec:Error-Propagation-Analysis}}

In this section, we focus on analyzing the EP phenomenon induced by
the IC in a recursive LocTrack system. First, we will present the
recursive expression of the EFIM. After this, we will examine the
convergence and asymptotic behavior of EP. In this section, all quantities
derived under recursive systems will be marked with a tilde (e.g.,
$\tilde{\mathbf{J}}_{e,t}$) to distinguish them from those derived
under the batch systems. 

\subsection{Recursive Expression of EFIM }

 In a recursive tracking system, at time $t$, we aim to estimate
$\boldsymbol{u}_{t}=\{\boldsymbol{u}_{k,t}\}_{k=1}^{K}$ based on
the measurements received up to the current time $\{\boldsymbol{y}_{\tau}\}_{\tau=1}^{t}$,
where $\boldsymbol{y}_{\tau}=[\boldsymbol{y}_{\tau,1}^{T},\cdots,\boldsymbol{y}_{\tau,K}^{T}]^{T}$.
Therefore, the EFIM for the current location $\mathbf{\tilde{J}}_{e,t}(\boldsymbol{u}_{t})$
(simplified as $\mathbf{\tilde{J}}_{e,t}$) can be expressed as a
recursive update of the previous EFIM $\mathbf{\tilde{J}}_{e,t-1}(\boldsymbol{u}_{t-1})$
(simplified as $\mathbf{\tilde{J}}_{e,t-1}),$ integrating the additional
information provided by the current time. In the following, we first
present the recursive expression of the EFIM $\mathbf{\tilde{J}}_{e,t}$,
and then analyze its convergence behaviors, including convergence
condition and convergence point. Let $\mathbf{\boldsymbol{\Lambda}}_{PS,t}^{o}$
denote the off-diagonal blocks of $-\boldsymbol{\Lambda}_{PS,t}$
in \eqref{eq:Gamma_PS}. Denoting $\boldsymbol{\Lambda}_{D,t}=\mathrm{BlockDiag}[\boldsymbol{\Lambda}_{D,t,1},\cdots,\boldsymbol{\Lambda}_{D,t,K}]$,
$\mathbf{\boldsymbol{\Lambda}}_{PS,t}^{D}=\mathrm{BlockDiag}[\mathbf{\boldsymbol{\Lambda}}_{PS,t,1}^{D},\cdots,\mathbf{\boldsymbol{\Lambda}}_{PS,t,K}^{D}]$
and $\boldsymbol{\varGamma}_{t}=\mathrm{BlockDiag}[\boldsymbol{\varGamma}_{t,1},\cdots,\boldsymbol{\varGamma}_{t,K}]$
with $\boldsymbol{\varGamma}_{t,k}$ given in \eqref{eq:Tem_FIM_2x2},
respectively, we have the following lemma.
\begin{lem}
[\label{lem:Tr-EFIM}Recursive Expression of $\mathbf{\tilde{J}}_{e,t}$]
The EFIM $\mathbf{\tilde{J}}_{e,t}$ of $\boldsymbol{u}_{t}$ can
be written as 
\begin{align}
\mathbf{\tilde{J}}_{e,t} & =\mathbf{\tilde{D}}_{t}\mathbf{\tilde{E}}_{t},\label{eq:dec_EFIM-1}
\end{align}
where the NPI $\mathbf{\tilde{D}}_{t}$ of $\boldsymbol{u}_{t}$ is
given by
\[
\mathbf{\tilde{D}}_{t}=\boldsymbol{\Lambda}_{D,t}+\mathbf{\boldsymbol{\Lambda}}_{PS,t}^{D}+\boldsymbol{\varGamma}_{t-1},
\]
with $\boldsymbol{\varGamma}_{0}=\boldsymbol{0}$. The EoC $\mathbf{\tilde{E}}_{t}$
is a function of $\mathbf{\tilde{J}}_{e,t-1}$, i.e.,
\begin{align}
\mathbf{\tilde{E}}_{t} & =\mathbf{I}-\mathbf{\tilde{D}}_{t}^{-1}\left[\underbrace{\mathbf{\boldsymbol{\Lambda}}_{PS,t}^{o}}_{\mathrm{Spatial\ IC}}+\underbrace{\boldsymbol{\varGamma}_{t-1}\left(\mathbf{\tilde{J}}_{e,t-1}+\boldsymbol{\varGamma}_{t-1}\right)^{-1}\boldsymbol{\varGamma}_{t-1}}_{\triangleq\mathbf{G}_{t-1}:\mathrm{Temporal\ IC}}\right].\label{eq:=000020Group_EoC}
\end{align}
\end{lem}
\begin{IEEEproof}
See Appendix \ref{subsec:Pf-Tr-EFIM}. 
\end{IEEEproof}
Similar to the batch system, the EFIM $\mathbf{\tilde{J}}_{e,t}$
in a recursive system can be decomposed into product form with NPI
$\mathbf{\tilde{D}}_{t}$ and EoC $\mathbf{\tilde{E}}_{t}$. NPI $\mathbf{\tilde{D}}_{t}$
consists of three components: information from measurements $\boldsymbol{\Lambda}_{D,t}$,
information from spatial neighbors $\mathbf{\boldsymbol{\Lambda}}_{PS,t}^{D}$,
and information from previous temporal neighbors $\boldsymbol{\varGamma}_{t-1}$,
under the assumption that no IC is presented. Notably, in comparison
to the NPI in the batch system, it can be shown that $\mathbf{\tilde{D}}_{t}\preceq\mathbf{D}_{t}$,
as the recursive system only captures temporal information from the
previous time step, rather than future time steps, leading to less
available temporal information.

The EoC $\mathbf{\tilde{E}}_{t}$ satisfies the condition $\mathbf{0}\prec\mathbf{\tilde{E}}_{t}\preceq\mathbf{I}$.
Both spatial IC and temporal IC can affect the efficiency. When the
information provided by the previous time is perfect, i.e., $\mathbf{\tilde{J}}_{e,t-1}\rightarrow\infty$
and $\mathbf{G}_{t-1}\rightarrow\mathbf{0}$, and the information
from the spatial neighbors is also perfect, i.e., $\mathbf{\boldsymbol{\Lambda}}_{PS,t}^{o}\rightarrow\mathbf{0}$,
we have $\mathbf{\tilde{E}}_{t}\rightarrow\mathbf{I}$. This indicates
no IC is present and full EoC is achieved. Note that $\mathbf{G}_{t-1}$
in \eqref{eq:=000020Group_EoC} quantifies the information loss induced
by the IC in the temporal domain. Specifically, the location estimation
error at the previous time can influence the current estimation, and
the impact of this error can propagate over time. Such EP phenomenon
can potentially jeopardize the overall localization performance. Therefore,
it is crucial to understand the stability of the EP in LocTrack systems,
as well as to examine the factors that influence it. Through theoretical
analysis, it is demonstrated that the well-conditioned input information
from measurements and appropriate STP prior can mitigate the EP phenomenon.
\begin{lem}
[Convergence condition of recursive systems\label{lem:Conv_cond-1}]The
NPI from current time $t$ satisfies the following condition:
\begin{align}
\underbrace{\boldsymbol{\Lambda}_{D,t}+\mathbf{\boldsymbol{\Lambda}}_{PS,t}^{D}+\boldsymbol{\varGamma}_{t-1}}_{\mathrm{NPI:}\tilde{\mathbf{D}_{t}}} & \succeq\mathbf{\tilde{J}}_{e,t-1}+\underbrace{\mathbf{\boldsymbol{\Lambda}}_{PS,t}^{o}}_{\mathrm{Spatial\ IC}}\nonumber \\
 & +\underbrace{\boldsymbol{\varGamma}_{t-1}\left(\mathbf{\tilde{J}}_{e,t-1}+\boldsymbol{\varGamma}_{t-1}\right)^{-1}\boldsymbol{\varGamma}_{t-1}}_{\mathrm{Temporal\ IC}:\mathbf{G}_{t-1}},\label{eq:Conv_cond-1}
\end{align}
then EP phenomenon can be eliminated as time goes up, i.e., $\mathbf{\tilde{J}}_{e,t}\succeq\mathbf{\tilde{J}}_{e,t-1}$.
\end{lem}
\begin{IEEEproof}
This Lemma can be obtained by rearranging terms straightforwardly.
\end{IEEEproof}
The convergence condition in \eqref{eq:Conv_cond-1} means that EFIM
obtained by NPI at current time should be able to compensate the spatiotemporal
IC, i.e., $\mathbf{\boldsymbol{\Lambda}}_{PS,t}^{o}+\mathbf{G}_{t-1}$,
to guarantee a decreased LocTrack error.  Therefore, to ensure a
converged LocTrack performance, high-quality NPI, proper STP prior,
or effective measurement policy is important for practical algorithm
designs. In most cases, system inputs can be considered to be stationary
over time. Assume that $\boldsymbol{\Lambda}_{D,t}+\mathbf{\boldsymbol{\Lambda}}_{PS,t}^{D}-\mathbf{\boldsymbol{\Lambda}}_{PS,t}^{o}$
and $\boldsymbol{\varGamma}_{t-1}$ in \eqref{eq:Conv_cond-1} remain
constant over time, and are denoted as $\mathbf{M}$ and $\mathbf{T}$,
respectively. Then, the convergence condition \eqref{eq:Conv_cond-1}
can be rewritten as
\begin{equation}
\mathbf{M}\succeq\left(\mathbf{\tilde{J}}_{e,t-1}^{-1}\mathbf{T}\mathbf{\tilde{J}}_{e,t-1}^{-1}+\mathbf{\tilde{J}}_{e,t-1}^{-1}\right)^{-1}\stackrel{\triangle}{=}\mathbf{\tilde{L}}_{t}.\label{eq:=000020Group_user_cond}
\end{equation}

\begin{figure}[t]
\centering{}\includegraphics[width=3in]{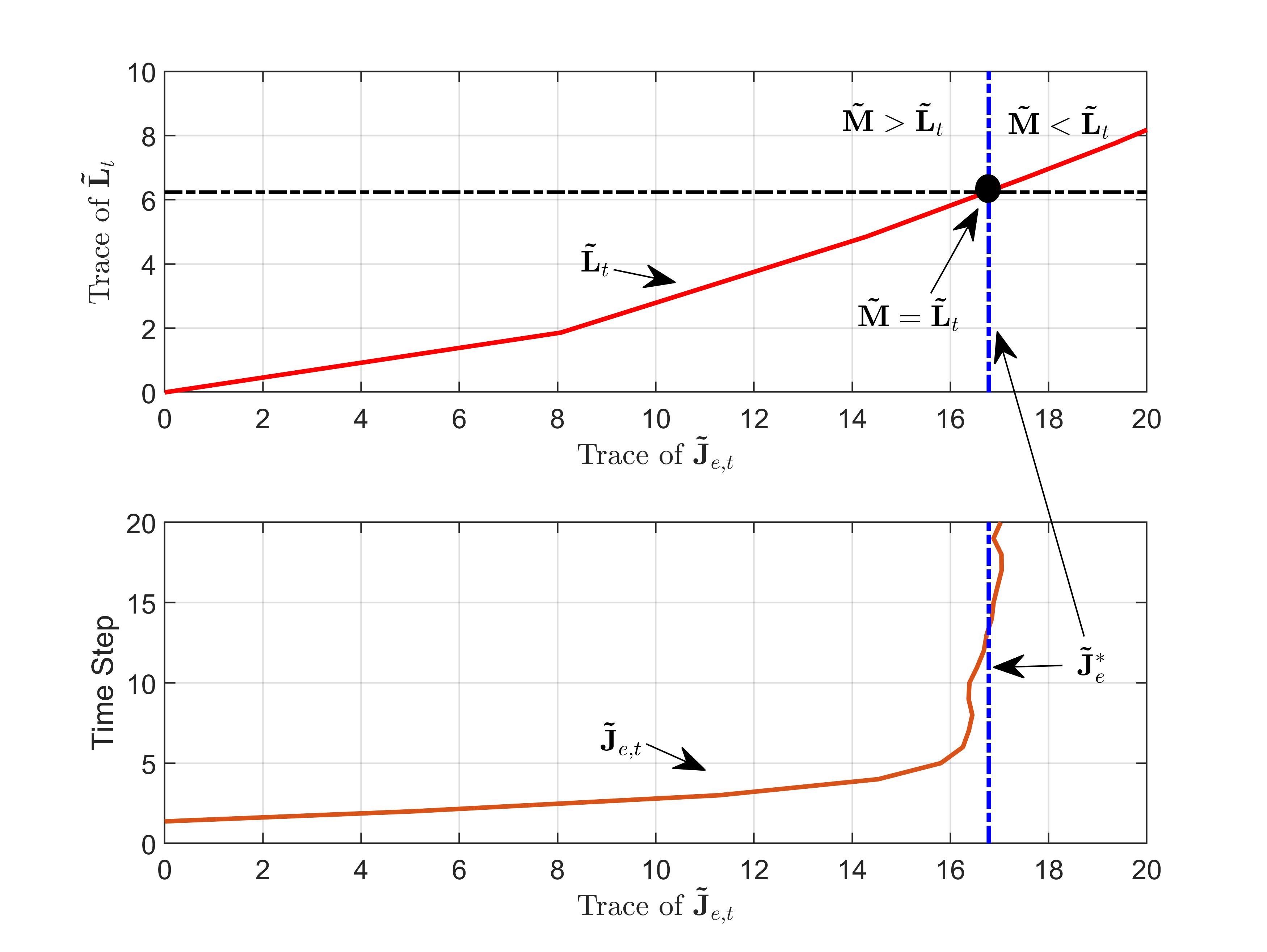}\caption{Illustration of convergence condition and convergence point.\protect\label{fig:EP_conv_cond}}
\end{figure}

The right hand side (RHS) can be interpreted as the temporal information
loss in the LocTrack due to the uncertainty in the previous location
estimate and the uncertainty in temporal position transition. As shown
in \eqref{eq:=000020Group_user_cond}, the RHS increases with $\mathbf{\tilde{J}}_{e,t-1}$,
which guarantees that $\mathbf{\tilde{J}}_{e,t}$ can converge to
a stationary point, i.e., $\mathbf{\tilde{J}}_{e,t}\rightarrow\mathbf{\tilde{J}}_{e}^{\star}$
. In particular, at the beginning of the LocTrack, there is very limited
prior information, resulting in a very small $\mathbf{\tilde{J}}_{e,t}$
and consequently a high localization error. Smaller $\mathbf{\tilde{J}}_{e,t}$
produces a reduced RHS, making it easier to satisfy condition \eqref{eq:=000020Group_user_cond}
and thus resulting in an increased $\mathbf{\tilde{J}}_{e,t+1}$ and
a decreased localization error. Therefore, as long as \eqref{eq:=000020Group_user_cond}
is satisfied, $\mathbf{\tilde{J}}_{e,t}$ will continue to increase.
When $\mathbf{\tilde{J}}_{e,t}$ reaches a level where the RHS of
\eqref{eq:=000020Group_user_cond} exceeds the LHS, the convergence
condition will be violated and $\mathbf{\tilde{J}}_{e,t+1}$ will
be smaller than $\mathbf{\tilde{J}}_{e,t}$. This decrease in $\mathbf{\tilde{J}}_{e,t+1}$
reduces the RHS, making the condition \eqref{eq:=000020Group_user_cond}
to be satisfied in the next time step. Consequently, the LocTrack
process converges to a stationary point, at which the LHS and RHS
of \eqref{eq:=000020Group_user_cond} are exactly balanced, as illustrated
in Fig. \ref{fig:EP_conv_cond}. The stationary point $\mathbf{\tilde{J}}_{e}^{\star}$
will be given in the following corollary.

\begin{cor}
[Convergence point of recursive systems\label{lem:Conv_point-1}]
In a sufficient long time, $\mathbf{\tilde{J}}_{e,t}$ will converge
to a stationary point $\mathbf{\tilde{J}}_{e}^{\star}$, which is
given by 
\begin{align}
\mathbf{\tilde{J}}_{e}^{\star} & =\left(\frac{1}{2}\mathbf{T}^{-\frac{1}{2}}\left(\mathbf{I}+4\mathbf{T}^{\frac{1}{2}}\mathbf{M}^{-1}\mathbf{T}^{\frac{1}{2}}\right)^{\frac{1}{2}}\mathbf{T}^{-\frac{1}{2}}-\frac{1}{2}\mathbf{T}^{-1}\right)^{-1}.\label{eq:Conv_point-1}
\end{align}
\end{cor}
\begin{IEEEproof}
See Appendix \ref{subsec:pf-conv-point}.
\end{IEEEproof}
From \eqref{eq:Conv_point-1}, the convergence point is jointly determined
by the measurements and STPs. Hence, carefully designing the LocTrack
system based on \eqref{eq:Conv_point-1} is important to guarantee
a low tracking error.
\begin{rem}
[Robustness in the LocTrack system\label{lem:Conv_rob}] We have
discussed the convergence behaviors of LocTrack systems with the constant
system inputs. However, the signal interference or hardware fault
can induce an unexpectable input and damage the estimation accuracy
dramatically. It is interesting that LocTrack system exhibits great
robustness towards the sudden failure. Suppose there is a tracking
failure at the time $\tilde{t}$, which leads to a small $\mathbf{\tilde{J}}_{e,\tilde{t}}$.
Similar to the initial stage of LocTrack, small $\mathbf{\tilde{J}}_{e,\tilde{t}}$
makes \eqref{eq:=000020Group_user_cond} easily to be satisfied. The
LocTrack system enters into the next steady-state cycle and reaches
the steady state eventually.
\end{rem}

\subsection{Recursive Expression of EFIM for Each User}

Following similar approach, we can derive the recursive form of $\mathbf{\tilde{J}}_{e,t,k}$.
\begin{lem}
[\label{lem:Tr-EFIM-each=000020user}Recursive Expression of $\mathbf{\tilde{J}}_{e,t,k}$]
The EFIM of the $k$-th user's position $\boldsymbol{u}_{t,k}$ can
be calculated by $\mathbf{\tilde{J}}_{e,t,k}=\left[\mathbf{\tilde{J}}_{e,t}\right]_{k,k}^{-1}$,
which is given by
\begin{align}
\mathbf{\tilde{J}}_{e,t,k} & =\mathbf{\tilde{D}}_{t,k}\mathbf{\tilde{E}}_{t,k},\label{eq:dec_EFIM-each-user}
\end{align}
where the NPI $\mathbf{\tilde{D}}_{t,k}$ is given by 
\begin{equation}
\mathbf{\tilde{D}}_{t,k}=\boldsymbol{\Lambda}_{D,t,k}+\boldsymbol{\Xi}_{k,k}^{t}+\boldsymbol{\varGamma}_{t-1,k}.\label{eq:=000020D_tk}
\end{equation}
The EoC $\mathbf{\tilde{E}}_{t,k}$ is given by
\begin{align}
\mathbf{\tilde{E}}_{t,k} & =\left[\mathbf{I}+\sum_{n=1}^{\infty}\left(\mathbf{\tilde{D}}_{t}^{-1}\left(\mathbf{\boldsymbol{\Lambda}}_{PS,t}^{o}+\mathbf{G}_{t-1}\right)\right)_{\gamma,\gamma}^{n}\right]^{-1},\label{eq:EoC_k}
\end{align}
where $\gamma=(t-1)K+k.$
\end{lem}
\begin{IEEEproof}
Similar to the proof of Lemma \ref{lem:Dec_EFIM}, which is omitted
here.
\end{IEEEproof}
Due to the complex expression of $\mathbf{\tilde{E}}_{t,k}$, it is
difficult to analyze the convergence behavior of $\mathbf{\tilde{J}}_{e,t,k}$
directly. In the next subsection, we will examine the convergence
behavior of $\mathbf{\tilde{J}}_{e,t,k}$ in three typical asymptotic
LocTrack scenarios.

\subsection{Asymptotic Discussion\protect\label{subsec:Asymptotic-Discussion-for}}

In practical LocTrack systems, the asymptotic performance provides
baselines to balance the performance gain and the IC in EP, and helps
us to demonstrate the impact of system parameters on EP behavior.
To simplify the analysis, we assume the spatiotemporal EFIM matrices
are in the extreme conditions, i.e., $\mathbf{\boldsymbol{\Lambda}}_{PS,t}\overset{\triangle}{=}\mathbf{\boldsymbol{\Lambda}}_{PS}^{Asy}\rightarrow\mathbf{0}/\infty$
or $\boldsymbol{\varGamma}_{t}\overset{\triangle}{=}\boldsymbol{\varGamma}^{Asy}\rightarrow\infty$.
Note that when $\boldsymbol{\varGamma}^{Asy}\rightarrow\mathbf{0}$,
the EFIM reduces to a simple case with only spatial correlations,
for which the analysis can be referenced in \cite{ICC_paper}.

\subsubsection{Impact of Small Spatial Correlation $\mathbf{\boldsymbol{\Lambda}}_{PS}^{Asy}\rightarrow\mathbf{0}$}

Based on \eqref{eq:=000020D_tk} and \eqref{eq:EoC_k}, when $\mathbf{\boldsymbol{\Lambda}}_{PS}^{Asy}\rightarrow\mathbf{0}$,
we can simplify $\mathbf{\tilde{D}}_{t,k}$ and $\mathbf{\tilde{E}}_{t,k}$
as 
\begin{align}
 & \lim_{\mathbf{\boldsymbol{\Lambda}}_{PS}^{Asy}\rightarrow\mathbf{0}}\mathbf{\tilde{D}}_{t,k}=\boldsymbol{\Lambda}_{D,t,k}+\boldsymbol{\varGamma}_{t-1,k},\label{eq:As->0}\\
 & \lim_{\mathbf{\boldsymbol{\Lambda}}_{PS}^{Asy}\rightarrow\mathbf{0}}\mathbf{\tilde{E}}_{t,k}\nonumber \\
 & =\mathbf{I}-\mathbf{\tilde{D}}_{t,k}^{-1}\underbrace{\boldsymbol{\varGamma}_{t-1,k}\left(\mathbf{\tilde{J}}_{e,t-1,k}+\boldsymbol{\varGamma}_{t-1,k}\right)^{-1}\boldsymbol{\varGamma}_{t-1,k}}_{\mathrm{Temporal\ IC}}.\label{eq:As->0=000020EOC}
\end{align}
Then we can obtain the convergence condition of $\mathbf{\tilde{J}}_{e,t,k}\succeq\mathbf{\tilde{J}}_{e,t-1,k}$
as follows:
\begin{align}
\boldsymbol{\Lambda}_{D,t,k}+\boldsymbol{\varGamma}_{t-1,k}\succeq & \mathbf{\tilde{J}}_{e,t-1,k}\nonumber \\
- & \boldsymbol{\varGamma}_{t-1,k}(\mathbf{\tilde{J}}_{e,t-1,k}+\boldsymbol{\varGamma}_{t-1,k})^{-1}\boldsymbol{\varGamma}_{t-1,k}.\label{eq:Cond_As->0}
\end{align}
The convergence state is given by
\begin{equation}
\begin{array}{cl}
 & \lim_{\,t\rightarrow\infty,\mathbf{\boldsymbol{\Lambda}}_{PS}^{Asy}\rightarrow\mathbf{0}}\mathbf{\tilde{J}}_{e,t,k}\\
 & =\left(\frac{1}{2}\mathbf{T}_{k}^{-\frac{1}{2}}\left(\mathbf{I}+4\mathbf{T}_{k}^{\frac{1}{2}}\mathbf{M}_{k}^{-1}\mathbf{T}_{k}^{\frac{1}{2}}\right)\mathbf{T}_{k}^{-\frac{1}{2}}-\frac{1}{2}\mathbf{T}_{k}^{-1}\right)^{-1},
\end{array}\label{eq:EP_pont_AS0}
\end{equation}
where $\mathbf{T}_{k}$ and $\mathbf{M}_{k}$ are the $(k,k)$-th
block submatrices of $\mathbf{T}=\boldsymbol{\varGamma}_{t-1}$ and
$\mathbf{M}=\boldsymbol{\Lambda}_{D,t}$ under constant input assumptions.
\begin{rem}
When the spatial correlation $\mathbf{\boldsymbol{\Lambda}}_{PS}^{Asy}$
goes to zero, the positions of users are uncorrelated with each other.
The analysis of localization performance can be decoupled across users.
Only temporal correlation can impact the LocTrack efficiency and EP
performance.
\end{rem}

\subsubsection{Impact of High Spatial Correlation $\mathbf{\boldsymbol{\Lambda}}_{PS}^{Asy}\rightarrow\boldsymbol{\infty}$}

Based on \eqref{eq:=000020D_tk} and \eqref{eq:EoC_k}, when $\mathbf{\boldsymbol{\Lambda}}_{PS}^{Asy}\rightarrow\mathbf{\mathbf{\infty}}$,
we can find that $\mathbf{\tilde{D}}_{t,k}\rightarrow\boldsymbol{\infty}$
and $\mathbf{\tilde{E}}_{t,k}\rightarrow\mathbf{0}$ by the following
Lemma.
\begin{lem}
[Asymptotic $\mathbf{\tilde{E}}_{t,k}$ when $\mathbf{\boldsymbol{\Lambda}}_{PS}^{Asy}\rightarrow\boldsymbol{\infty}$\label{lem:rank=000020of=000020E_tk}]
When the spatial correlation $\mathbf{\boldsymbol{\Lambda}}_{PS}^{Asy}$
tends to infinity, we have 
\[
\lim_{\mathbf{\boldsymbol{\Lambda}}_{PS}^{Asy}\rightarrow\mathbf{\mathbf{\mathbf{\infty}}}}\mathbf{\tilde{E}}_{t,k}=\left[\mathbf{I}+\sum_{n=1}^{\infty}\left(\mathbf{\tilde{D}}_{t}^{-1}\mathbf{\boldsymbol{\Lambda}}_{PS,t}^{o}\right)_{\gamma,\gamma}^{n}\right]^{-1}\rightarrow\mathbf{0}.
\]
\end{lem}
\begin{IEEEproof}
See Appendix \ref{subsec:Pf-rank_Etk}.
\end{IEEEproof}
To obtain the convergence condition of EP in this scenario and derive
the convergence point of $\mathbf{\tilde{J}}_{e,t,k}$ when $\mathbf{\boldsymbol{\Lambda}}_{PS}^{Asy}\rightarrow\mathbf{\mathbf{\infty}}$,
we have the following equivalent expression of $\mathbf{\tilde{J}}_{e,t,k}$.
\begin{lem}
[Equivalent expression of $\mathbf{\tilde{J}}_{e,t,k}$ when $\mathbf{\boldsymbol{\Lambda}}_{PS}^{Asy}\rightarrow\boldsymbol{\infty}$\label{lem:Eq_As-0}]
When the spatial correlation $\mathbf{\boldsymbol{\Lambda}}_{PS}^{Asy}$
tends to infinity, the EFIM $\mathbf{\tilde{J}}_{e,t,k}$ is equivalent
to 
\begin{align}
\lim_{\mathbf{\boldsymbol{\Lambda}}_{PS}^{Asy}\rightarrow\boldsymbol{\infty}}\mathbf{\tilde{J}}_{e,t,k} & =\boldsymbol{\Lambda}_{s}^{t}+\boldsymbol{\varGamma}_{t-1,s}\label{eq:=000020equiv_Jetk}\\
 & -\boldsymbol{\varGamma}_{t-1,s}\left(\mathbf{\tilde{J}}_{e,t-1,k}+\boldsymbol{\varGamma}_{t-1,s}\right)^{-1}\boldsymbol{\varGamma}_{t-1,s},\nonumber 
\end{align}
where $\boldsymbol{\varGamma}_{t-1,s}$ and $\boldsymbol{\Lambda}_{s}^{t}$
are given by $\boldsymbol{\varGamma}_{t-1,s}=\sum_{k}\boldsymbol{\varGamma}_{t-1,k}$
and $\boldsymbol{\Lambda}_{s}^{t}=\sum_{k}\boldsymbol{\Lambda}_{D,t,k}$,
respectively. 
\end{lem}
\begin{IEEEproof}
See Appendix \ref{subsec:Pf-AS_lemma}.
\end{IEEEproof}
Based on above lemma, the EP convergence condition can be formulated
as 
\begin{align*}
\boldsymbol{\Lambda}_{D,t,s}+\boldsymbol{\varGamma}_{t-1,s} & \succeq\mathbf{\tilde{J}}_{e,t-1,k}\\
 & -\boldsymbol{\varGamma}_{t-1,s}(\mathbf{\tilde{J}}_{e,t-1,k}+\boldsymbol{\varGamma}_{t-1,s})^{-1}\boldsymbol{\varGamma}_{t-1,s}.
\end{align*}
Compared with \eqref{eq:Cond_As->0}, it shows that the spatial correlation
is beneficial for EP convergence, since when strong special correlation
is present, the convergence condition of EP is more easily satisfied,
facilitating the mitigation of EP. Denoting $\mathbf{M}_{s}=\sum_{k}\mathbf{M}_{k}$
and $\mathbf{T}_{s}=\sum_{k}\mathbf{T}_{k}$ where $\mathbf{M}_{k}$
and $\mathbf{T}_{k}$ are the $(k,k)$-th block submatrices of $\mathbf{M}=\boldsymbol{\Lambda}_{D,t}$
and $\mathbf{T}=\boldsymbol{\varGamma}_{t-1}$, respectively, under
constant input assumptions, the convergence point can be given by
\begin{equation}
\begin{array}{cl}
 & \lim_{\,t\rightarrow\infty,\mathbf{\boldsymbol{\Lambda}}_{PS}^{Asy}\rightarrow\boldsymbol{\infty}}\widetilde{\mathbf{J}}_{e,t,k}\\
 & =\left(\frac{1}{2}\mathbf{T}_{s}^{-\frac{1}{2}}\left(\mathbf{I}+4\mathbf{T}_{s}^{\frac{1}{2}}\mathbf{M}_{s}^{-1}\mathbf{T}_{s}^{\frac{1}{2}}\right)\mathbf{T}_{s}^{-\frac{1}{2}}-\frac{1}{2}\mathbf{T}_{s}^{-1}\right)^{-1}.
\end{array}\label{eq:Asy_EP_lar_sp_point}
\end{equation}

\begin{rem}
With strong spatial correlations, multiple users become transparent
to each other, allowing positional information obtained from measurements
and temporal correlations to be superimposed effectively across users,
thereby enhancing the overall localization accuracy. As spatial correlation
approaches infinity, the IC between users also tends to infinity.
The position error of one user can directly affect the position estimation
of the neighboring users. From the perspective of RWM, the flow of
positional information among users forms a closed loop, leading to
limited effective information flow to the BS. Consequently, the EoC
approaches zero.
\end{rem}

\subsubsection{Impact of High Temporal Correlation $\boldsymbol{\varGamma}^{Asy}\rightarrow\boldsymbol{\infty}$}

Finally, we examine the asymptotic limits when the temporal correlation
is at the high level. When $\boldsymbol{\varGamma}^{Asy}\rightarrow\boldsymbol{\infty}$,
we have $\lim_{\boldsymbol{\varGamma}_{t-1}\rightarrow\boldsymbol{\infty}}\mathbf{\tilde{D}}_{t,k}\rightarrow\boldsymbol{\infty}$
based on \eqref{eq:=000020D_tk} and $\mathbf{\tilde{E}}_{t,k}\rightarrow\mathbf{0}$
by the following Lemma.
\begin{lem}
[Asymptotic $\mathbf{\tilde{E}}_{t,k}$ when $\boldsymbol{\varGamma}^{Asy}\rightarrow\boldsymbol{\infty}$\label{lem:lar_tem=000020of=000020E_tk-1}]
When the temporal correlation $\boldsymbol{\varGamma}^{Asy}$ tends
to infinity, we have 
\[
\lim_{\boldsymbol{\varGamma}^{Asy}\rightarrow\mathbf{\mathbf{\mathbf{\infty}}}}\mathbf{\tilde{E}}_{t,k}=\left[\mathbf{I}+\sum_{n=1}^{\infty}\left(\mathbf{\tilde{D}}_{t}^{-1}\mathbf{G}_{t-1}\right)_{\gamma,\gamma}^{n}\right]^{-1}\rightarrow\mathbf{0}.
\]
\end{lem}
\begin{IEEEproof}
This proof is similar to the proof of Lemma \ref{lem:rank=000020of=000020E_tk}
which is omitted here.
\end{IEEEproof}
To obtain the convergence condition of EP in this scenario and derive
the convergence point of $\mathbf{\tilde{J}}_{e,t,k}$ when $\boldsymbol{\varGamma}^{Asy}\rightarrow\mathbf{\mathbf{\infty}}$,
we have the following equivalent expression of $\mathbf{\tilde{J}}_{e,t,k}$.
\begin{lem}
[Equivalent expression of $\mathbf{\tilde{J}}_{e,t,k}$ when $\boldsymbol{\varGamma}^{Asy}\rightarrow\boldsymbol{\infty}$\label{lem:Eq_At_inf}]
When the temporal correlation $\boldsymbol{\varGamma}^{Asy}$ tends
to infinity, the EFIM $\mathbf{\tilde{J}}_{e,t,k}$ is equivalent
to 
\begin{align}
\lim_{\boldsymbol{\varGamma}^{Asy}\rightarrow\mathbf{\mathbf{\mathbf{\infty}}}}\mathbf{\tilde{J}}_{e,t,k} & =\sum_{\tau=1}^{t}\left(\boldsymbol{\Lambda}_{D,\tau,k}+\boldsymbol{\Xi}_{k,k}^{\tau}\right)\nonumber \\
 & \times\left[\mathbf{I}+\sum_{n=1}^{\infty}\left(\mathbf{\tilde{D}}_{t}^{-1}\sum_{\tau=1}^{t}\mathbf{\boldsymbol{\Lambda}}_{PS,\tau}^{o}\right)_{\gamma,\gamma}^{n}\right]^{-1}.\label{eq:=000020equiv_Jetk-1}
\end{align}
\end{lem}
\begin{IEEEproof}
See Appendix \ref{subsec:Pf-At_inf}.
\end{IEEEproof}
When system inputs are stable over time, $\boldsymbol{\Lambda}_{D,t,k}$,
$\boldsymbol{\Xi}_{k,k}^{t}$ and $\mathbf{\boldsymbol{\Lambda}}_{PS,\tau}^{o}$
in \eqref{eq:=000020equiv_Jetk-1} are constant. Denoting $\boldsymbol{\Lambda}_{D,t,k}+\boldsymbol{\Xi}_{k,k}^{t}=\mathbf{M}_{k}$
and $\left[\mathbf{I}+\sum_{n=1}^{\infty}\left(\mathbf{\tilde{D}}_{t}^{-1}\sum_{\tau=1}^{t}\mathbf{\boldsymbol{\Lambda}}_{PS,\tau}^{o}\right)_{\gamma,\gamma}^{n}\right]^{-1}=\mathbf{C}_{k}$,
the convergence condition $\mathbf{\tilde{J}}_{e,t,k}\succeq\mathbf{\tilde{J}}_{e,t-1,k}$
becomes 
\begin{equation}
t\mathbf{M}_{k}\succeq\mathbf{\tilde{J}}_{e,t-1,k}+t\mathbf{M}_{k}\left(\mathbf{I}-\mathbf{C}_{k}\right).\label{eq:=000020inf_conv_cond}
\end{equation}
Since $\mathbf{\tilde{J}}_{e,t-1,k}=(t-1)\mathbf{M}_{k}\mathbf{C}_{k}$,
convergence condition in \eqref{eq:=000020inf_conv_cond} is automatically
satisfied. Consequently, the EFIM $\mathbf{\tilde{J}}_{e,t,k}$ will
go to infinity and BCRB will converge to zero when the time step $t$
goes to infinity.
\begin{rem}
With strong temporal correlation, each user accumulates positional
information from observations and spatial correlations at each time
step. When the accumulated information approaches infinity, the localization
error tends toward zero. As temporal correlation goes to infinity,
this indicates that the IC across time also tends to infinity. The
position error of previous time slot can directly affect the position
estimation of the current time, leading to strong positional information
flow over time with limited effective information flow to the BS.
Consequently, the EoC approaches zero.
\end{rem}

\section{Simulation Results \protect\label{sec:Simulation-Results}}

\subsection{System Setting}

\begin{figure}[t]
\centering{}\includegraphics[width=3in]{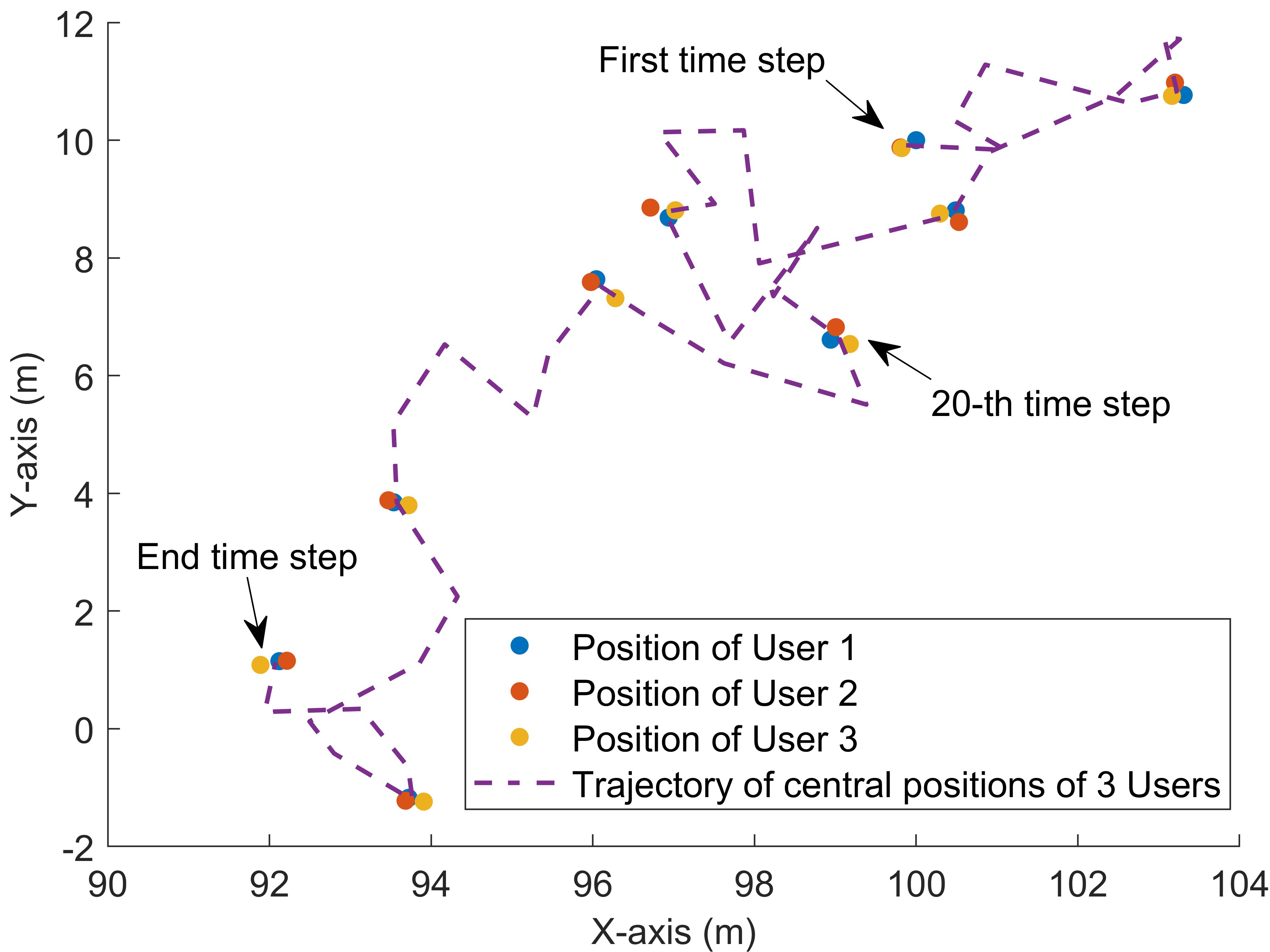}\caption{The movement trajectories of 3 users over 40 time steps. \protect\label{fig:Sim-Trajectory} }
\end{figure}

In this section, we present the simulation results to verify theoretical
conclusions obtained in this paper. Consider $K=3$ users in the $R=4$
RIS-assisted LocTrack system with spatiotemporal correlations. The
temporal correlations for each user are captured by Gaussian distribution
as in \eqref{eq:Temproal_corr} with $\sigma_{t,k}^{-2}=\sigma_{tem}^{-2}=10,t\in[1,T-1],k\in\mathcal{K}$.
We consider pairwise spatial correlations, where the inter-user distance
are captured by Gaussian distribution as in \eqref{eq:Pedestrian}
with $\sigma_{t,ij}^{-2}=\sigma_{spa}^{-2}=10,t\in\mathcal{T},i,j\in\mathcal{K},i\neq j$.
The other system settings are listed as follows: antenna numbers $N_{B}=64$,
$N_{r}=32$, carriers frequency $f_{c}=28$ GHz, noise variance $\sigma_{t}^{-2}=1$,
$t\in\mathcal{T}$, transmit power 10 dBm, LoS path loss factor $\alpha=-2.08$
and multi-path factor $\kappa_{BR}=\kappa_{RU}=20$ dB \cite{tRIS-tr1}.
Positions of BS and $R=4$ RISs are located at $\boldsymbol{b}=[0,0]^{T}$m
and $\boldsymbol{r}_{i}=[80,25+5i]^{T}$m, $\forall i\in\mathcal{R}$.
Positions of users at the initial time step $t=1$ are located at
$\boldsymbol{u}_{1,1}=[100,10]^{T}$m, $\boldsymbol{u}_{1,2}=[110,10]^{T}$m
and $\boldsymbol{u}_{1,3}=[105,10+5\sqrt{3}]^{T}$m, respectively,
which randomly move from the initial positions following the predefined
spatiotemporal correlation model. One set of typical users' trajectories
is shown in Fig. \ref{fig:Sim-Trajectory}. All simulation results
are the averages of 2000 Mont-Carlos experiments. 

\subsection{Convergence Behavior of EP}

\begin{figure}[t]
\centering{}\includegraphics[width=3in]{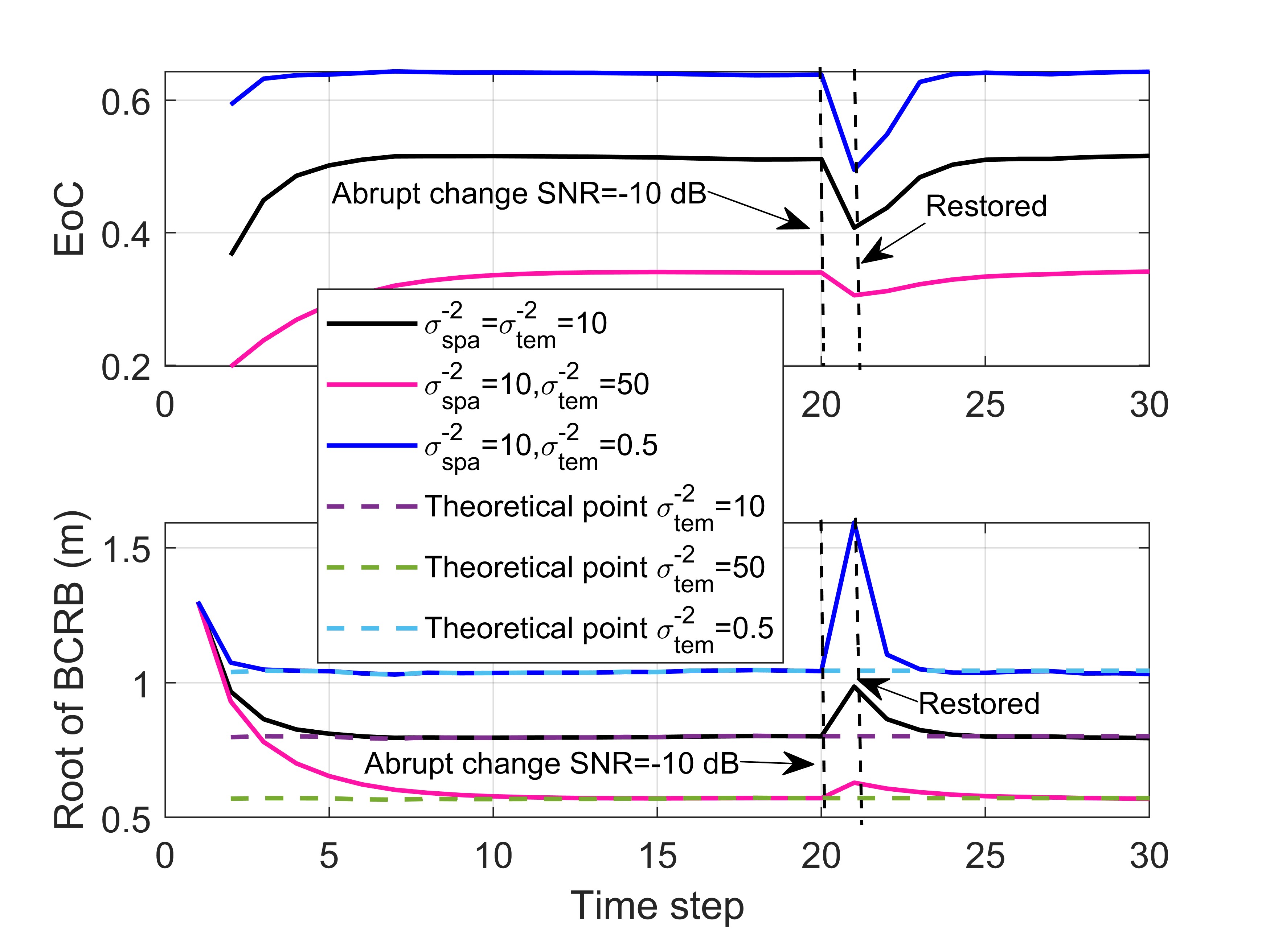}\caption{EP converge behaviors under different levels of temporal correlations.
\protect\label{fig:EP_con_tem}}
\end{figure}

We then investigate the convergence behaviors of averaged EoC, i.e.,
$\frac{1}{2K}\mathrm{Tr}(\tilde{\mathbf{E}}_{t}$) in \eqref{eq:=000020Group_EoC},
and root of BCRB, i.e., $\sqrt{\frac{1}{2K}\mathrm{Tr}(\mathbf{\tilde{J}}_{e,t}^{-1})}$
in \eqref{eq:dec_EFIM-1} under different levels of temporal correlations
in a recursive LocTrack system. The theoretical value of converged
BCRB can be calculated as $BCRB^{\star}=\frac{1}{2K}\mathrm{Tr}\left[(\mathbf{\tilde{J}}_{e}^{\star})^{-1}\right]$,
where $\mathbf{\tilde{J}}_{e}^{\star}$ is given in \eqref{eq:Conv_point-1}.
As shown in Fig. \ref{fig:EP_con_tem}, the trends of BCRB align well
with the theoretical predictions (dash lines), and the converged values
match the theoretical results. Furthermore, a comparison across different
correlation levels reveals that strong correlation enhances the BCRB
performance. However, the resulting strong IC may hinder efficient
information utilization, leading to lower EoC. Therefore, there is
a trade-off performance between BCRB and EoC. Additionally, the level
of correlation influences the convergence speed of both the EoC and
the BCRB. Specifically, higher correlation levels decelerate the convergence
of these performance metrics.

To test the robustness, we simulate an abrupt change by reducing the
SNR to -10 dB at time step 21, restoring at time step 23. The results
show that the LocTrack system is resilient to abrupt changes, exhibiting
strong robustness, particular under scenarios with strong correlation
priors.

\subsection{Asymptotic Behavior of EP}

\begin{figure}[t]
\centering{}\includegraphics[width=3in]{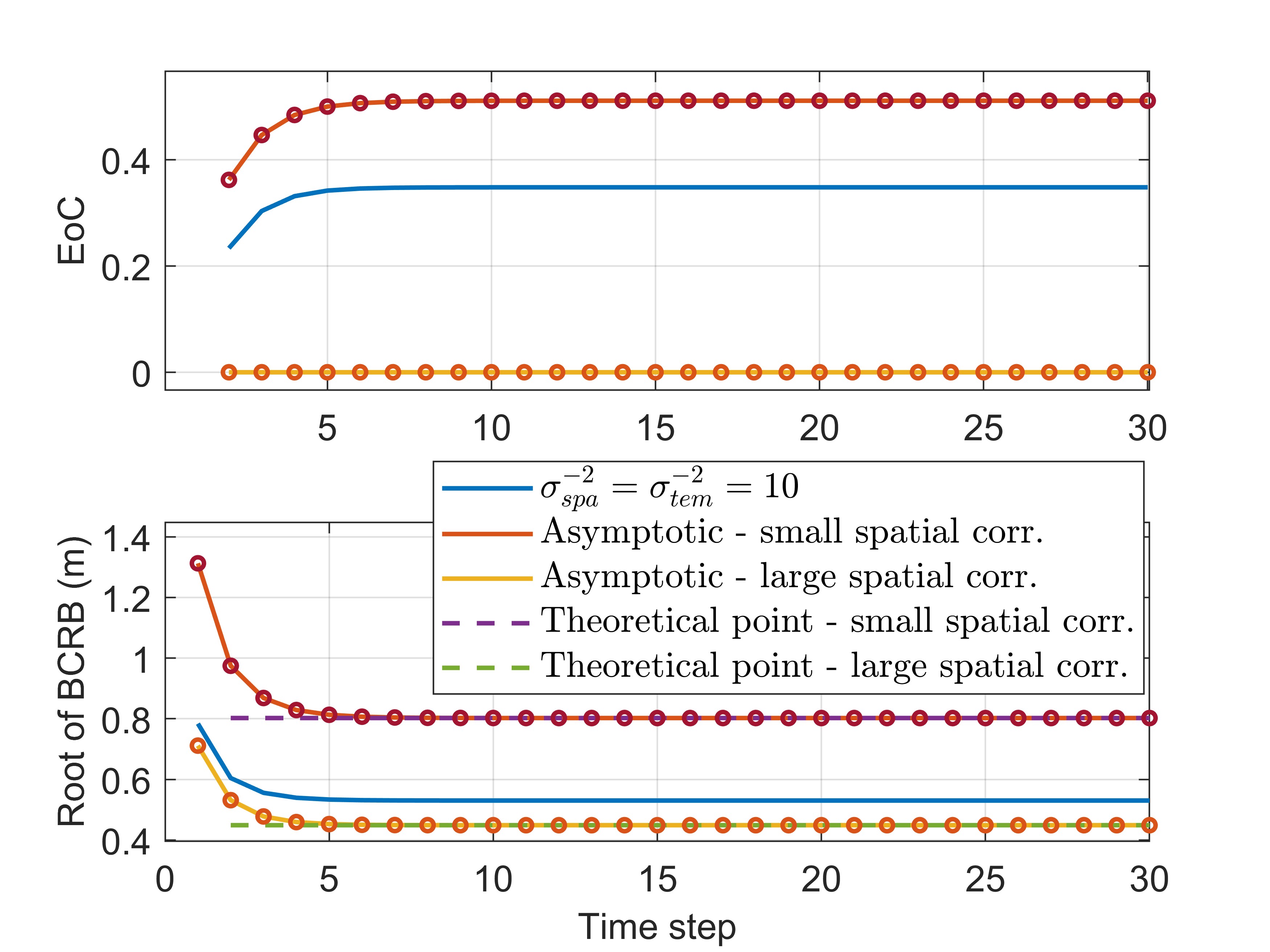}\caption{Asymptotic EP behaviors in $\mathbf{\boldsymbol{\Lambda}}_{PS}^{Asy}\rightarrow\boldsymbol{0}$
and $\mathbf{\boldsymbol{\Lambda}}_{PS}^{Asy}\rightarrow\mathbf{\infty}$.
Solid lines are derived from \eqref{eq:=000020Group_EoC} and \eqref{eq:dec_EFIM-1}.
Dash lines represent the theoretical convergence points of BCRB calculated
by \eqref{eq:EP_pont_AS0} (zero spatial correlation), \eqref{eq:Conv_point-1}
(general spatial correlation) and \eqref{eq:Asy_EP_lar_sp_point}
(infinite spatial correlation). The small circles represent the asymptotic
EoC and BCRB values derived from the theoretical analysis. \protect\label{fig:Asy_EP_spa}}
\end{figure}

Finally, we investigate the asymptotic behaviors of averaged EoC,
i.e., $\frac{1}{2K}\mathrm{Tr}(\tilde{\mathbf{E}}_{t}$) in \eqref{eq:=000020Group_EoC},
and root of BCRB, i.e., $\sqrt{\frac{1}{2K}\mathrm{Tr}(\mathbf{\tilde{J}}_{e,t}^{-1})}$
in \eqref{eq:dec_EFIM-1} under four asymptotic scenarios. We set
$\sigma_{spa}^{-2}=$1e-3, $\sigma_{spa}^{-2}=$1e3, $\sigma_{tem}^{-2}=$1e-3
and $\sigma_{tem}^{-2}=$1e3 to represent the cases of $\mathbf{\boldsymbol{\Lambda}}_{PS}^{Asy}\rightarrow\boldsymbol{0}$,
$\mathbf{\boldsymbol{\Lambda}}_{PS}^{Asy}\rightarrow\infty$, $\boldsymbol{\varGamma}^{Asy}\rightarrow\boldsymbol{0}$
and $\boldsymbol{\varGamma}^{Asy}\rightarrow\boldsymbol{\infty}$,
respectively. As illustrated in Fig. \ref{fig:Asy_EP_spa}, when the
spatial correlation is extremely high, such that the inter-user position
uncertainties vanish, the strong coupling between users leads to rapid
propagation of measurement information among users, thereby significantly
reducing the EoC to zero. Instead, the transparency among users allows
the positional information to be shared and accumulated, thereby enhancing
the LocTrack accuracy. When the spatial correlation is negligible
such that multiple users can be considered uncorrelated in spatial
domain, only temporal IC contributes to the EoC, resulting in a higher
EoC. The lack of spatial priors degrades LocTrack accuracy compared
to strong spatial correlations.

\begin{figure}[t]
\centering{}\includegraphics[width=3in]{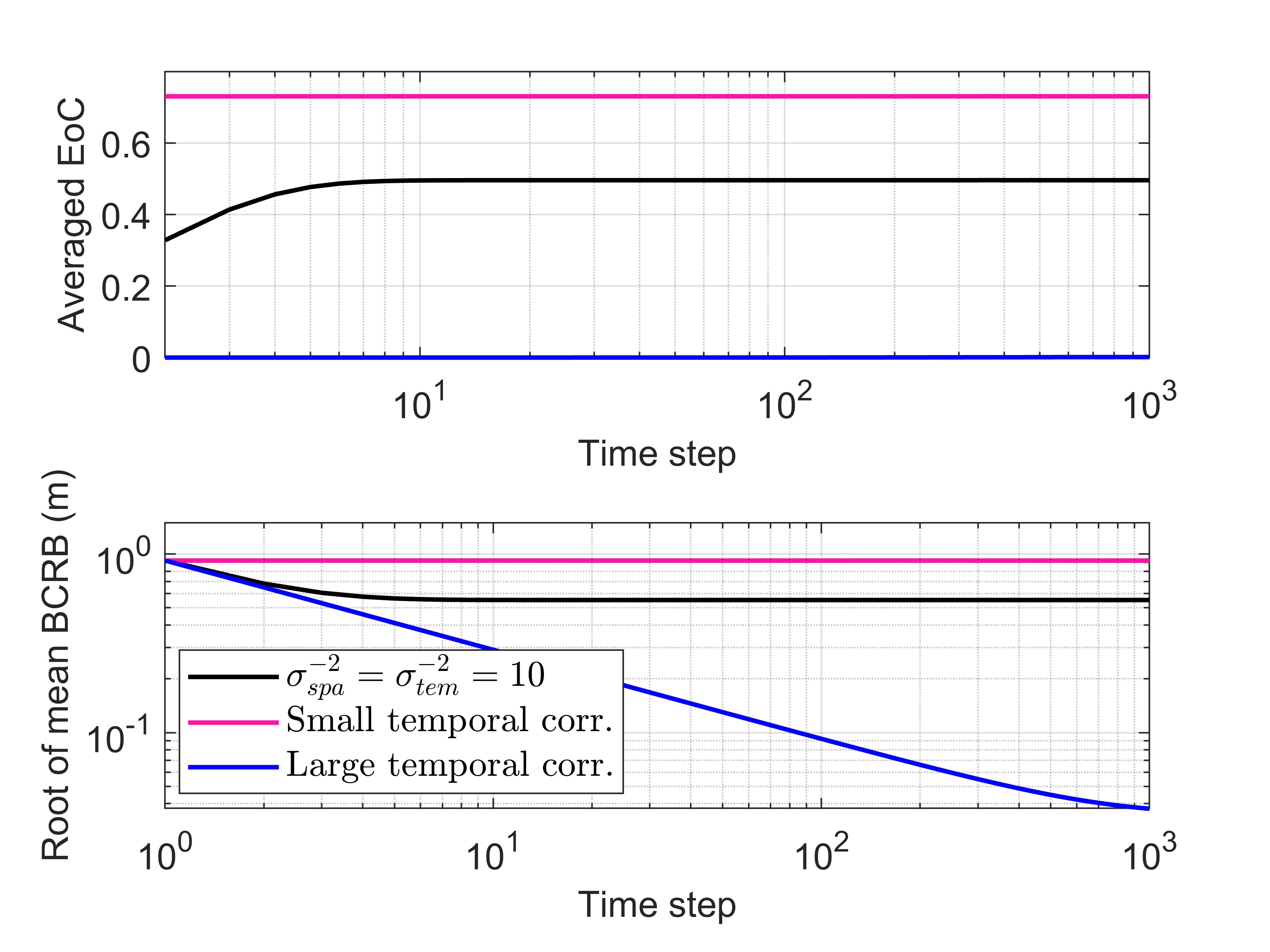}\caption{Asymptotic EP behaviors in $\boldsymbol{\varGamma}^{Asy}\rightarrow\boldsymbol{0}$
and $\boldsymbol{\varGamma}^{Asy}\rightarrow\boldsymbol{\infty}$
with $T=1000$. Solid lines are derived from \eqref{eq:=000020Group_EoC}
and \eqref{eq:dec_EFIM-1}. \protect\label{fig:Asy_EP_tem}}
\end{figure}

As shown in Fig. \ref{fig:Asy_EP_tem}, when the temporal correlation
tends to infinity such that the position transitions over time becomes
deterministic, spatial IC can propagate over time transparently, leading
to a zero EoC. Conversely, this temporal transparency allows the positional
information to accumulate over time. As time goes to infinity, the
BCRB asymptotically approaches zero with log-linear convergence rate.
When temporal correlation is weak such that users are uncorrelated
over time, only current spatial correlation can affect the EoC, resulting
in a higher and steady EoC over time. In this case, less positional
information can be obtained, leading to reduced LocTrack accuracy.

\section{Conclusion \protect\label{sec:Conclusion}}

In this paper, we establish a theoretical foundation for analyzing
the fundamental relationships between localization performance and
STP among users' positions in a multi-RIS assisted multi-user localization
and tracking system. We first propose a novel metric EoC by decomposing
the EFIM, which indicates the efficiency of information utilization
in LocTrack system. Then we analyze the IC phenomenon in a batch LocTrack
system and provide a graph interpretation of EoC by RWM theory to
reveal the correlational position information routing. We also analyze
the EP phenomenon in a recursive LocTrack system and characterize
its convergence behaviors and asymptotic behaviors based on EoC metric
to unveil the EP principle. Simulation results are provided to verify
the correctness of our proposed theoretical foundations.

\appendix

\appendices{}

\subsection{Proof of Lemma \ref{lem:EFIM} \protect\label{subsec:Proof=000020of=000020EFIM}}

Consider a mapping from $\boldsymbol{\eta}$ to another parameter
vector $\boldsymbol{\omega}$, where $\boldsymbol{\omega}=[\boldsymbol{\theta}^{T},\boldsymbol{\xi}^{T}]^{T}$
and $\boldsymbol{\theta}=\left[\boldsymbol{\theta}_{1,1}^{T},\ldots,\boldsymbol{\theta}_{t,k}^{T},\ldots,\boldsymbol{\theta}_{T,K}^{T}\right]^{T}$
with 
\begin{equation}
\boldsymbol{\theta}_{t,k}^{T}=[\theta_{1,1,1}^{RU},\rho_{1,1,1}^{RU},\ldots,\theta_{R,t,k}^{RU},\rho_{R,t,k}^{RU}]^{T}.\label{eq:Theta_tk}
\end{equation}
 The FIM of $\boldsymbol{\eta}$ can be calculated as
\begin{equation}
\mathbf{J}_{\boldsymbol{\eta}}=\left[\begin{array}{cc}
\mathbf{T}_{\boldsymbol{u}}\boldsymbol{\Lambda}_{\boldsymbol{\theta}}\mathbf{T}_{\boldsymbol{u}}^{T} & \mathbf{T}_{\boldsymbol{u}}\boldsymbol{\Lambda}_{\boldsymbol{\xi}\boldsymbol{\theta}}^{T}\\
\boldsymbol{\Lambda_{\boldsymbol{\xi}\boldsymbol{\theta}}}\mathbf{T}_{\boldsymbol{u}}^{T} & \boldsymbol{\Lambda_{\boldsymbol{\xi}}}
\end{array}\right]+\left[\begin{array}{cc}
\mathbf{\boldsymbol{\Lambda}}_{P} & \mathbf{0}\\
\mathbf{0} & \mathbf{0}
\end{array}\right].\label{eq:J_eta}
\end{equation}

Using the Schur's complement for parameter $\boldsymbol{\theta}$,
we have $\mathbf{J}_{e}=\mathbf{T}_{\boldsymbol{u}}\left(\boldsymbol{\Lambda}_{\boldsymbol{\theta}}-\boldsymbol{\Lambda}_{\boldsymbol{\xi}\boldsymbol{\theta}}^{T}\boldsymbol{\Lambda}_{\boldsymbol{\xi}}^{-1}\boldsymbol{\Lambda}_{\boldsymbol{\xi}\boldsymbol{\theta}}\right)\mathbf{T}_{\boldsymbol{u}}^{T}$$+\mathbf{\boldsymbol{\Lambda}}_{P}\triangleq\mathrm{\boldsymbol{\Lambda}_{D}}+\mathbf{\boldsymbol{\Lambda}}_{P}$.
The submatrices in \eqref{eq:J_eta} are given as follows. The Jacobian
matrix for the transformation from $\boldsymbol{u}$ to $\boldsymbol{\theta}$
is given by $\mathbf{T}_{\boldsymbol{u}}=\frac{\partial\boldsymbol{\theta}}{\partial\boldsymbol{u}},$
which is a block diagonal matrix. $\boldsymbol{\Lambda}_{\boldsymbol{\theta}}=-\mathbb{E}_{\boldsymbol{y,\eta}}\left[\frac{\partial^{2}\ln p\left(\boldsymbol{y}\mid\boldsymbol{\eta}\right)}{\partial\boldsymbol{\theta}\partial\boldsymbol{\theta}^{T}}\right]$,
$\boldsymbol{\Lambda_{\boldsymbol{\xi}}}=-\mathbb{E}_{\boldsymbol{y,\eta}}\left[\frac{\partial^{2}\ln p\left(\boldsymbol{y}\mid\boldsymbol{\eta}\right)}{\partial\boldsymbol{\boldsymbol{\xi}}\partial\boldsymbol{\boldsymbol{\xi}}^{T}}\right]$
and $\boldsymbol{\Lambda_{\boldsymbol{\xi}\boldsymbol{\theta}}}=-\mathbb{E}_{\boldsymbol{y,\eta}}\left[\frac{\partial^{2}\ln p\left(\boldsymbol{y}\mid\boldsymbol{\eta}\right)}{\partial\boldsymbol{\boldsymbol{\xi}}\partial\boldsymbol{\theta}^{T}}\right]$,
which are also block-diagonal matrices calculated through received
signal model in \eqref{eq:y_k}. Therefore, $\boldsymbol{\Lambda}_{D}$
in \eqref{eq:Lambda_D} is also a block-diagonal matrix.

Since prior $p\left(\boldsymbol{u}\right)$ in \eqref{eq:prior_dis}
is in log-linear form, by $\mathbf{\boldsymbol{\Lambda}}_{P}=-\mathbb{E}_{\boldsymbol{u}}\left[\frac{\partial^{2}}{\partial\boldsymbol{u}\partial\boldsymbol{u}^{T}}\ln p\left(\boldsymbol{u}\right)\right]$,
we have 
\begin{equation}
\mathbf{\boldsymbol{\Lambda}}_{P}=\left[\begin{array}{ccccc}
\mathbf{\boldsymbol{\Lambda}}_{P}^{1,1} & \mathbf{\boldsymbol{\Lambda}}_{P}^{1,2} & \mathbf{0} & \cdots & \mathbf{0}\\
\mathbf{\boldsymbol{\Lambda}}_{P}^{2,1} & \mathbf{\boldsymbol{\Lambda}}_{P}^{2,2} & \mathbf{\boldsymbol{\Lambda}}_{P}^{2,3} & \cdots & \mathbf{0}\\
\vdots & \vdots & \vdots & \ddots & \vdots\\
\mathbf{0} & \mathbf{0} & \mathbf{0} & \cdots & \mathbf{\boldsymbol{\Lambda}}_{P}^{T,T}
\end{array}\right],\label{eq:Gamma_P}
\end{equation}
with $\mathbf{\boldsymbol{\Lambda}}_{P}^{i,j}=-\mathbb{E}_{\boldsymbol{u}}\left[\frac{\partial^{2}}{\partial\boldsymbol{u}_{i}\partial\boldsymbol{u}_{j}^{T}}\ln p\left(\boldsymbol{u}\right)\right]$.
Due to the Markovian property of $p(\boldsymbol{u})$ in temporal
domain, we have $\boldsymbol{\Lambda}_{p}^{i,j}=0,|i-j|\geq2$. For
any continuous and differentiable functions of distance $f(d_{t}^{i,j})$
and $g(d_{t-1,t}^{i})$, where $d_{t}^{i,j}=\left\Vert \boldsymbol{u}_{t,i}-\boldsymbol{u}_{t,j}\right\Vert $
and $d_{t-1,t}^{i}=\left\Vert \boldsymbol{u}_{t-1,i}-\boldsymbol{u}_{t,i}\right\Vert $,
$\forall i,j\in\mathcal{K},$ we have
\[
\frac{\partial^{2}f(d_{t}^{i,j})}{\partial\boldsymbol{u}_{t,i}\partial\boldsymbol{u}_{t,j}}=\frac{\partial^{2}f(d_{t}^{i,j})}{\partial\boldsymbol{u}_{t,j}\partial\boldsymbol{u}_{t,i}},\frac{\partial^{2}g(d_{t-1,t}^{i})}{\partial\boldsymbol{u}_{t-1,i}\partial\boldsymbol{u}_{t,i}}=\frac{\partial^{2}g(d_{t-1,t}^{i})}{\partial\boldsymbol{u}_{t,i}\partial\boldsymbol{u}_{t-1,i}},
\]
\[
\frac{\partial^{2}f(d_{t}^{i,j})}{\partial\boldsymbol{u}_{t,i}\partial\boldsymbol{u}_{t,i}}=-\frac{\partial^{2}f(d_{t}^{i,j})}{\partial\boldsymbol{u}_{t,i}\partial\boldsymbol{u}_{t,j}},\frac{\partial^{2}g(d_{t-1,t}^{i})}{\partial\boldsymbol{u}_{t,i}\partial\boldsymbol{u}_{t,i}}=-\frac{\partial^{2}g(d_{t-1,t}^{i})}{\partial\boldsymbol{u}_{t-1,i}\partial\boldsymbol{u}_{t,i}}.
\]
Since the pairwise potential functions $\varphi(\boldsymbol{u}_{t,i},\boldsymbol{u}_{t,j})$
and temporal transition probability function $p(\boldsymbol{u}_{t+1,k}|\boldsymbol{u}_{t,k})$
in \eqref{eq:Temproal_corr} can be written as functions of inter-position
distances, they both satisfy the above properties. Therefore, $\mathbf{\boldsymbol{\Lambda}}_{P}$
in \eqref{eq:Gamma_P} has the following properties:
\begin{itemize}
\item $\mathbf{\boldsymbol{\Lambda}}_{P}=\mathbf{\boldsymbol{\Lambda}}_{PS}+\mathbf{\boldsymbol{\Lambda}}_{PT}$.
\item $\mathbf{\boldsymbol{\Lambda}}_{PS}=\mathrm{BlockDiag}\left[\mathbf{\boldsymbol{\Lambda}}_{PS,1},\ldots,\mathbf{\boldsymbol{\Lambda}}_{PS,T}^ {}\right]$
with 
\begin{equation}
\mathbf{\boldsymbol{\Lambda}}_{PS,t}=\left[\begin{array}{ccc}
\boldsymbol{\Xi}_{1,1}^{t} & \cdots & \boldsymbol{\Xi}_{1,K}^{t}\\
\vdots & \ddots & \vdots\\
\boldsymbol{\Xi}_{K,1}^{t} & \cdots & \boldsymbol{\Xi}_{K,K}^{t}
\end{array}\right],\label{eq:Gamma_PS}
\end{equation}
where
\begin{equation}
\boldsymbol{\Xi}_{i,j}^{t}=\begin{cases}
\frac{-2\partial^{2}}{\partial\boldsymbol{u}_{t,i}\partial\boldsymbol{u}_{t.j}^{T}}\ln\varphi(\boldsymbol{u}_{t,i},\boldsymbol{u}_{t,j}) & (i,j)\in E,\\
\frac{-2\partial^{2}}{\partial\boldsymbol{u}_{t,i}\partial\boldsymbol{u}_{t.i}^{T}}\sum_{i,j}\ln\varphi(\boldsymbol{u}_{t,i},\boldsymbol{u}_{t,j}) & i=j,\\
\mathbf{0} & (i,j)\notin E,
\end{cases}\label{eq:Gamma_PS_detail}
\end{equation}
Therefore, we have $\left[\mathbf{\boldsymbol{\Lambda}}_{PS}\right]_{i,i}=-\sum_{j\in\mathcal{I}_{i}}\left[\mathbf{\boldsymbol{\Lambda}}_{PS}\right]_{i,j}$.
\item $\mathbf{\boldsymbol{\Lambda}}_{PT}$ has the following form: 
\begin{equation}
\mathbf{\boldsymbol{\Lambda}}_{PT}=\left[\begin{array}{cccc}
\boldsymbol{\varGamma}_{1} & -\boldsymbol{\varGamma}_{1} & \cdots & \mathbf{0}\\
-\boldsymbol{\varGamma}_{1} & \boldsymbol{\varGamma}_{1}+\boldsymbol{\varGamma}_{2} & \cdots & \mathbf{0}\\
\vdots & \vdots & \ddots & \vdots\\
\mathbf{0} & \mathbf{0} & -\boldsymbol{\varGamma}_{T-1} & \boldsymbol{\varGamma}_{T-1}
\end{array}\right],\label{eq:Gamma_PT}
\end{equation}
where $\boldsymbol{\varGamma}_{t}=\mathrm{BlockDiag}\left[\boldsymbol{\varGamma}_{t,1},\cdots,\boldsymbol{\varGamma}_{t,K}\right]$
is a block-diagonal matrix with 
\begin{equation}
\boldsymbol{\varGamma}_{t,k}=\frac{\partial^{2}}{\partial\boldsymbol{u}_{t,k}\partial\boldsymbol{u}_{t+1,k}^{T}}\ln p(\boldsymbol{u}_{t+1,k}|\boldsymbol{u}_{t,k}).\label{eq:Tem_FIM_2x2}
\end{equation}
Therefore, we have $\left[\mathbf{\boldsymbol{\Lambda}}_{PT}\right]_{i,i}=$$-\sum_{j\in\mathcal{I}_{i}}\left[\mathbf{\boldsymbol{\Lambda}}_{PT}\right]_{i,j}=-\left[\mathbf{\boldsymbol{\Lambda}}_{PT}\right]_{i,i-K}-\left[\mathbf{\boldsymbol{\Lambda}}_{PT}\right]_{i,i+K},$with
$\left[\mathbf{\boldsymbol{\Lambda}}_{PT}\right]_{i,j}=\boldsymbol{0},$
if $j<0$ or $j>K$.
\end{itemize}

\subsection{Proof of Lemma \ref{lem:Dec_EFIM}\protect\label{subsec:Pf-Lemma-Dec}}

Taking matrix inverse for both sides of \eqref{eq:difference}, we
have $\mathbf{J}_{e}^{-1}=\left(\mathbf{I}-\mathbf{D}^{-1}\mathbf{A}\right)^{-1}\mathbf{D}^{-1}.$
Denoting $\mathbf{Q}=\mathbf{D}^{-1}\mathbf{A},$ and expanding $\left(\mathbf{I}-\mathbf{Q}\right)^{-1}$
using power series, the inverse of EFIM can be expressed as 
\[
\mathbf{J}_{e}^{-1}=\left(\mathbf{I}+\sum_{n=1}^{\infty}\text{\ensuremath{\mathbf{Q}}}^{n}\right)\mathbf{D}^{-1},
\]
which converges as long as $\mathbf{J}_{e}^{-1}$ exists. Hence,
\begin{equation}
\mathbf{J}_{e,t,k}^{-1}=\left[\mathbf{J}_{e}^{-1}\right]_{\gamma,\gamma}=\left(\mathbf{I}+\sum_{n=1}^{\infty}\left[\text{\ensuremath{\mathbf{Q}}}^{n}\right]_{\gamma,\gamma}\right)\mathbf{D}_{t,k}^{-1}.\label{eq:proof_inv}
\end{equation}
with $\gamma=(t-1)K+k$ for brevity. Taking the inverse, we have
\begin{equation}
\mathbf{J}_{e,t,k}=\mathbf{D}_{t,k}\left(\mathbf{I}+\sum_{n=1}^{\infty}\left[\text{\ensuremath{\mathbf{Q}}}^{n}\right]_{\gamma,\gamma}\right)^{-1}.\label{eq:proof_Q}
\end{equation}

To facilitate the RWM interpretation, we augment $\text{\ensuremath{\mathbf{Q}}}$
to generate a transition matrix $\mathbf{P}$ as defined in \eqref{eq:Absorb}.
Then we have $\left[\mathbf{P}^{n}\right]_{\gamma,\gamma}=\text{\ensuremath{\left[\mathbf{Q}^{n}\right]_{\gamma,\gamma}}}$.
Combined with \eqref{eq:proof_Q}, we can get \eqref{eq:dec_EFIM}.
Note that $\text{\ensuremath{\mathbf{J}_{e,t,k}}}$ is a real symmetric
matrix with $\mathbf{J}_{e,t,k}\preceq\mathbf{D}_{t,k}$, which implies
$\boldsymbol{\Delta}_{t,k}\succeq\mathbf{0}$. Then the proof is completed.

\subsection{Proof of Lemma \ref{lem:Tr-EFIM}\protect\label{subsec:Pf-Tr-EFIM}}

The EFIM of $\boldsymbol{u}^{(t)}=\{\boldsymbol{u}_{\tau}\}_{\tau=1}^{t}$
is given by
\[
\widetilde{\mathbf{J}}_{e}^{(t)}=\left[\begin{array}{c|c}
\widetilde{\mathbf{J}}_{e}^{(t-1)}+\left[\begin{array}{ccc}
\mathbf{0} & \cdots & \mathbf{0}\\
\vdots & \ddots & \mathbf{\vdots}\\
\mathbf{0} & \cdots & \boldsymbol{\varGamma}_{t-1}
\end{array}\right] & \left[\begin{array}{c}
\mathbf{0}\\
\mathbf{\vdots}\\
-\boldsymbol{\varGamma}_{t-1}
\end{array}\right]\\
\hline \left[\begin{array}{ccc}
\mathbf{0} & \cdots & -\boldsymbol{\varGamma}_{t-1}\end{array}\right] & \mathbf{\tilde{D}}_{t}-\mathbf{\boldsymbol{\Lambda}}_{PS,t}^{o}
\end{array}\right].
\]
Based on the Schur's complement operation, we have
\[
\tilde{\mathbf{J}}_{e,t}=\mathbf{\tilde{D}}_{t}-\mathbf{\boldsymbol{\Lambda}}_{PS,t}^{o}-\boldsymbol{\varGamma}_{t-1}\left(\tilde{\mathbf{J}}_{e,t-1}+\boldsymbol{\varGamma}_{t-1}\right)^{-1}\boldsymbol{\varGamma}_{t-1}.
\]
Then we can derive \eqref{eq:=000020Group_EoC} readily.

\subsection{Proof of Convergence Point in Corollary \ref{lem:Conv_point-1}\protect\label{subsec:pf-conv-point} }

The limits point in \eqref{eq:Conv_point-1} can be obtained by solving
equality $\tilde{\mathbf{J}}_{e,t}=\tilde{\mathbf{J}}_{e,t-1}$,
\begin{equation}
\mathbf{M}=\left(\mathbf{\tilde{J}}_{e,t-1}^{-1}\mathbf{T}\mathbf{\tilde{J}}_{e,t-1}^{-1}+\mathbf{\tilde{J}}_{e,t-1}^{-1}\right)^{-1}.\label{eq:pf_cond_eq}
\end{equation}
Define the limit point of $\mathbf{\tilde{J}}_{e,t}$ is $\mathbf{X}$,
\eqref{eq:pf_cond_eq} can be rewritten as 
\begin{equation}
\mathbf{M}+\mathbf{T}-\mathbf{T}\left(\mathbf{X}+\mathbf{T}\right)^{-1}\mathbf{T}=\mathbf{X}.\label{eq:pf_cond_eq2}
\end{equation}
This is a typical Recitati equation that hard to solve. One particular
solution can be obtain by 
\begin{equation}
\mathbf{X}^{-1}=\frac{1}{2}\mathbf{T}^{-\frac{1}{2}}\left(\mathbf{I}+4\mathbf{T}^{\frac{1}{2}}\mathbf{M}^{-1}\mathbf{T}^{\frac{1}{2}}\right)^{\frac{1}{2}}\mathbf{T}^{-\frac{1}{2}}-\frac{1}{2}\mathbf{T}^{-1}.\label{eq:pf_cond_eq3}
\end{equation}
It can be verified by substituting \eqref{eq:pf_cond_eq3} into \eqref{eq:pf_cond_eq2}
easily.

\subsection{Proof of Lemma \ref{lem:rank=000020of=000020E_tk} \protect\label{subsec:Pf-rank_Etk}}

The asymptotic expression of $\mathbf{\tilde{E}}_{t,k}$ can be obtained
by the fact of temporal IC $\mathbf{G}_{t-1}\preceq\boldsymbol{\varGamma}_{t-1}\preceq\mathbf{\infty}$,
due to $\mathbf{G}_{t-1}=\boldsymbol{\varGamma}_{t-1}\left(\mathbf{\tilde{J}}_{e,t-1}+\boldsymbol{\varGamma}_{t-1}\right)^{-1}\boldsymbol{\varGamma}_{t-1}\preceq\min\left\{ \boldsymbol{\varGamma}_{t-1},\mathbf{\tilde{J}}_{e,t-1}\right\} $
derived from matrix inverse lemma. Then we have 
\[
\mathbf{\tilde{E}}_{t,k}\rightarrow\left[\mathbf{I}+\sum_{n=1}^{\infty}\left(\mathbf{\tilde{D}}_{t}^{-1}\mathbf{\boldsymbol{\Lambda}}_{PS,t}^{o}\right)_{\gamma,\gamma}^{n}\right]^{-1},
\]
when $\mathbf{\boldsymbol{\Lambda}}_{PS}^{Asy}\rightarrow\boldsymbol{\infty}$.
Recall the RWM theory and the graph interpretation in Fig. \ref{fig:RWM},
$\mathbf{\tilde{E}}_{t,k}$ at time step $t$ can be equivalent to
$K$ users random walk without any temporal correlations. Thus the
PTPM based on \eqref{eq:transition} among spatial correlated users
is $-\mathbf{D}_{t,k}^{-1}\boldsymbol{\Xi}_{k,j}^{t}$ with $\sum_{j\in E}\mathbf{D}_{t,k}^{-1}\boldsymbol{\Xi}_{k,j}^{t}=\mathbf{I}$
where $(i,j)\in E$, while the PTPM between BS and itself is $\mathbf{D}_{t,k}^{-1}\boldsymbol{\Lambda}_{D,t,k}\rightarrow\mathbf{0}$
since $\mathbf{D}_{t,k}\rightarrow\boldsymbol{\infty}$. It implies
information trapped into spatial users, resulting $\mathbf{\tilde{E}}_{t,k}\rightarrow\mathbf{0}$.

\subsection{Proof of Lemma \ref{lem:Eq_As-0} \protect\label{subsec:Pf-AS_lemma}}

Let us focus on the first time step in tracking system,
\[
\widetilde{\mathbf{J}}_{e,1}=-\left[\begin{array}{cccc}
\mathbf{S}_{1} & \boldsymbol{\Xi}_{1,2} & \cdots & \mathbf{0}\\
\boldsymbol{\Xi}_{1,2} & \mathbf{S}_{2} & \ddots & \mathbf{0}\\
\vdots & \ddots & \ddots & \boldsymbol{\Xi}_{K-1,K}\\
\mathbf{0} & \mathbf{0} & \boldsymbol{\Xi}_{K-1,K} & \mathbf{S}_{K}
\end{array}\right],
\]
where $\mathbf{S}_{k}=\boldsymbol{\Lambda}_{D,k}+\boldsymbol{\Xi}_{k,k}$
and time index here is omitted for brevity. As $\mathbf{\boldsymbol{\Lambda}}_{PS}^{Asy}\rightarrow\boldsymbol{\infty}$,
we can obtain $\mathbf{\tilde{J}}_{e,1,K}$ by Schur's complement
similar as proof of Lemma \ref{subsec:Pf-Tr-EFIM}, i.e., 
\begin{align*}
\mathbf{\tilde{J}}_{e,K} & =\boldsymbol{\Lambda}_{D,K}+\left(\boldsymbol{\Xi}_{K-1,K}^{-1}+\left(\left[\widetilde{\mathbf{J}}_{e,1}\right]_{1:K-1,1:K-1}\right)^{-1}\right)^{-1}\\
 & \overset{\mathbf{\boldsymbol{\Lambda}}_{S}\rightarrow\boldsymbol{\infty}}{=}\boldsymbol{\Lambda}_{D,K}+\boldsymbol{\Lambda}_{D,K-1}+\mathbf{\tilde{J}}_{e,1}\mid_{k=1}^{K-1},
\end{align*}
where $\mathbf{\tilde{J}}_{e,1}\mid_{k=1}^{K-1}$ represents removing
$K$-th user from $\widetilde{\mathbf{J}}_{e,1}$. Then repeat the
above operation, we have property that $\lim_{\mathbf{\boldsymbol{\Lambda}}_{PS}^{Asy}\rightarrow\boldsymbol{\infty}}\mathbf{\tilde{J}}_{e,1,K}=\sum_{k=1}^{K}\boldsymbol{\Lambda}_{D,1,k}$.
Similar results can be extended to any user. Note that by the above
property, as $\mathbf{\boldsymbol{\Lambda}}_{PS}^{Asy}\rightarrow\boldsymbol{\infty}$,
$\mathbf{\tilde{J}}_{e,1,K}$ tends to identity matrix, and hence
we can obtain the equality in the Lemma for any time slot. Therefore,
we complete the proof.

\subsection{Proof of Lemma \ref{lem:Eq_At_inf} \protect\label{subsec:Pf-At_inf}}

Based on \eqref{eq:Conv_cond-1}, we have the following fact 

\begin{align*}
\lim_{\boldsymbol{\varGamma}_{t-1}\rightarrow\boldsymbol{\infty}}\left(\boldsymbol{\varGamma}_{t-1}-\mathbf{G}_{t-1}\right) & =\mathbf{\tilde{J}}_{e,t-1},
\end{align*}
indicating previous EFIM of $K$ users can be perfectly transferred
into the current time under strong temporal correlations. Rearrange
the block-diagonal submatrix as below
\begin{align}
\lim_{\boldsymbol{\varGamma}_{t-1}\rightarrow\boldsymbol{\infty}}\check{\mathbf{D}}_{t,k} & =\boldsymbol{\Lambda}_{D,t,k}+\boldsymbol{\Xi}_{k,k}^{t}+\left[\mathbf{\tilde{J}}_{e,t-1}\right]_{k,k}\nonumber \\
 & =\sum_{\tau=1}^{t}\left(\boldsymbol{\Lambda}_{D,\tau,k}+\boldsymbol{\Xi}_{k,k}^{\tau}\right),\label{eq:AT-inf-Dk}
\end{align}
then the EoC-like matrix can be rewritten as $\lim_{\boldsymbol{\varGamma}_{t-1}\rightarrow\boldsymbol{\infty}}\check{\mathbf{E}}_{t,k}=\left[\mathbf{I}+\sum_{n=1}^{\infty}\left(\mathbf{\tilde{D}}_{t}^{-1}\sum_{\tau=1}^{t}\mathbf{\boldsymbol{\Lambda}}_{PS,\tau}^{o}\right)_{\gamma,\gamma}^{n}\right]^{-1}$
with an equivalent EFIM expression, i.e., $\mathbf{\tilde{J}}_{e,t,k}=\check{\mathbf{D}}_{t,k}\check{\mathbf{E}}_{t,k}$.

\bibliographystyle{IEEEtran}
\bibliography{IEEEabrv,reference_abrv}

\end{document}